\documentclass[prd,preprint,tightenlines,showpacs,preprintnumbers,nofootinbib,eqsecnum,superscriptaddress]{revtex4}

 \usepackage[dvips,final]{graphicx}
  \usepackage{amssymb}
   \usepackage{amsmath}
    \usepackage{amsfonts}
     \usepackage{bm}% bold math
      \usepackage{makecell}
\newcommand{\mycites}[1]{Refs.~\cite{#1}}
\newcommand{\mycite}[1]{Ref.~\cite{#1}}
\newcommand{\vect}[1]{\boldsymbol{#1}}

\begin{document}
\begin{flushright}
DESY 14-120\\
July 2014
\end{flushright}
\vskip+5mm
\title{Conventional versus single-ladder-splitting contributions \\ 
to double parton scattering production \\
of two quarkonia, two Higgs bosons and $c \bar c c \bar c$. 
} 

\author{Jonathan R. Gaunt}
\email{jonathan.gaunt@desy.de}
\affiliation{Theory Group, Deutsches Elektronen-Synchrotron (DESY), D-22607 Hamburg, Germany}

\author{Rafa{\l} Maciu{\l}a}
\email{rafal.maciula@ifj.edu.pl}
\affiliation{Institute of Nuclear Physics PAN, PL-31-342 Cracow, Poland}

\author{Antoni~Szczurek}
\email{antoni.szczurek@ifj.edu.pl}
\affiliation{University of Rzesz\'ow, PL-35-959 Rzesz\'ow, Poland}
\affiliation{Institute of Nuclear Physics PAN, PL-31-342 Cracow, Poland} 

\date{\today}

\begin{abstract}
The double parton distributions (dPDF), both conventional and those
corresponding to parton splitting, are calculated and compared
for different two-parton combinations. The conventional and splitting dPDFs
have very similar shape in $x_1$ and $x_2$.
We make a first quantitative evaluation of the single-ladder-splitting contribution to
double parton scattering (DPS) production of two S- or P-wave quarkonia,
two Higgs bosons and $c \bar c c \bar c$.
The ratio of the single-ladder-splitting to conventional contributions is discussed
as a function of centre-of-mass energy, mass of the produced system
and other kinematical variables. Using a simple model for the dependence of
the conventional two-parton distribution on transverse parton separation (Gaussian
and independent of $x_i$ and scales), we find that 
the 2v1 contribution is as big as the 2v2 contribution 
discussed in recent years in the literature. 
This means that the phenomenological analyses of 
$\sigma_{eff}$ including only the conventional DPS mechanism have to be revised 
including explicitly the single-ladder-splitting contributions discussed here.
The differential distributions in rapidity and transverse momenta 
calculated for conventional and single-ladder-splitting DPS processes are however very similar
which causes their experimental separation to be rather
difficult, if not impossible.
The direct consequence of the existence of the two components (conventional and
splitting) is the energy and process dependence of the empirical parameter 
$\sigma_{eff}$. This is illustrated in our paper for the considered processes.

\end{abstract}

\pacs{12.38.Bx, 11.80.La, 14.40.Lb, 14.40.Pq, 14.80.Bn}
%12.38.Aw 	General properties of QCD (dynamics, confinement, etc.) 
%12.38.Bx 	Perturbative calculations
%11.80.La 	Multiple scattering
%14.40.Lb 	Charmed mesons (|C|>0, B=0) 
%14.40.Pq 	Heavy quarkonia
%14.80.Bn 	Standard-model Higgs bosons

\maketitle

%----------------------------
\section{Introduction}
%----------------------------

The LHC, as the highest energy collider ever available, is 
the best place to study double parton scattering (DPS) (or, 
more generally, multi-parton interactions, or MPI). This fact
triggered several recent theoretical studies of DPS.
The theoretical understanding of DPS is not yet complete.
Double parton distributions in the proton depending only on the 
longitudinal momentum fractions $x_1, x_2$ of the two partons and the 
corresponding scales $\mu_1^2,\mu_2^2$ were introduced long ago, 
and evolution equations for these quantities were derived \cite{K79, 
SSZ82,S2003,GS2010}. However, more recently \cite{Gaunt:2011xd, Gaunt:2012dd, Blok:2011bu, 
Manohar:2012jr, Manohar:2012pe, DS2011,DOS2012, RS2011} it was established that 
these quantities are not adequate to describe proton-proton DPS (although 
they may be used to describe the dominant contribution to the proton-heavy nucleus DPS 
process \cite{Cattaruzza:2004qb, Strikman:2001gz}). Rather, one should 
describe this process in terms of two-parton generalised parton distributions
(2pGPDs), which aside from the momentum fractions and scales of the two
partons also depend on the transverse impact parameter between the partons,
$b$. The work of \mycites{Gaunt:2011xd, Gaunt:2012dd, Blok:2011bu, RS2011}
involved considering low order Feynman diagrams and then generalising 
the findings to allow a resummation outwards from the hard process, whilst
that of \mycites{Manohar:2012jr, Manohar:2012pe, DS2011, DOS2012} was somewhat
more formal in nature and laid down some first steps towards a factorisation
proof for DPS (with \mycites{Manohar:2012jr, Manohar:2012pe} utilising the method
of soft collinear effective theory, and \mycites{DS2011, DOS2012} following the 
more traditional Collin-Soper-Sterman approach).

One important finding of the work in \mycites{Gaunt:2012dd, Blok:2011bu,
Manohar:2012pe, RS2011} was that there are (at least) two different 
types of contribution to the DPS cross section, which are accompanied
by different geometrical prefactors. One of these is the ``conventional''
or 2v2 contribution in which two separate ladders emerge from both
protons and interact in two separate hard interactions -- this one has
been well-known for a long time \cite{Politzer:1980me, Paver:1982yp} 
and is the one that is often considered in phenomenological analyses.
The other type of process is the ``perturbative ladder splitting''
or 2v1 contribution, which is similar to the 2v2 process except that
one proton initially provides one ladder, which later perturbatively
splits into two. The 2v1 contribution to the DPS cross section has not
received much attention in numerical studies, apart from one study \cite{BDFS2013}
that gives estimates of the size of the effect in four-jet, $\gamma$
+ 3j, $W^+jj$ and $W^+W^-$ production. There may also be a 1v1 
contribution to DPS in which there is a perturbative ladder splitting 
in both protons, although there is some controversy in the literature 
over whether this process should entirely be regarded as single 
parton scattering (SPS), or if there is a portion of it that can be 
regarded as DPS \cite{Cacciari:2009dp, Blok:2011bu, Gaunt:2011xd, 
DOS2012, Manohar:2012pe, RS2011}.

In \mycite{BDFS2013} a sizable effect of the 2v1 ladder splitting process
was observed for the processes studied there with a rather weak dependence 
on the kinematical variables. This indicates that the ladder splitting 
process may be important for other DPS processes studied at the LHC.
In this paper we will study the relative importance of the conventional
2v2 and ladder splitting 2v1 processes, for various processes whose 
production is dominated by gluon-gluon fusion. The representative 
examples are e.g. production of two S-wave ($\eta$) or P-wave ($\chi$) 
quarkonia, two Higgs bosons and double open charm.
The last process was studied recently by two of us
\cite{LMS2012,MS2013,HMS2014}. A cross section for the process was 
estimated and detailed comparison to experimental data obtained by 
the LHCb collaboration \cite{LHCb} was made. Even including 
higher-order corrections in the $k_t$-factorization approach some 
deficit of the cross section was observed \cite{HMS2014}, at least with the standard set of parameters. 
This deficit cannot be understood as due to leading-order 
single parton scattering $g g \to c \bar c c \bar c$ mechanism 
\cite{SS2012,HMS2014}, and it is interesting to investigate if it
can be at least partially due to the parton splitting contribution. 

In the following we shall quantify the splitting 2v1 contribution 
for these processes and discuss its influence on the so-called
effective cross section measured by comparison of the factorized 
model with experimental data. We generalize the formula for the 
total cross section from \mycite{Gaunt:2012dd} to the case of 
differential distributions. Since our focus is on the relative 
contribution from the 2v2 and 2v1 contributions, we will not 
consider any possible 1v1 contribution to DPS (the method by 
which one would calculate such a contribution within DPS is 
anyway unclear at the present moment). Effectively we are 
therefore following \mycites{Blok:2011bu, Gaunt:2011xd, Manohar:2012pe}
and just taking such 1v1 processes to be pure SPS.

%--------------------------------------------
\section{Sketch of the formalism}
%--------------------------------------------

In this section we present a sketch of the formalism used
to calculate the splitting 2v1 and nonsplitting 2v2 contributions
to double quarkonium (double Higgs boson) and $c \bar c c \bar c$ production.
Various notations for calculating these contributions have been used in the 
literature -- in the following we shall use the one from \mycite{Gaunt:2012dd}.

%-------------------------------------------------------------------
\subsection{DPS production of two quarkonia
and two Higgs bosons}  \label{sec:DPSsetup}
%-------------------------------------------------------------------

\begin{figure}[!h]
\includegraphics[width=5cm]{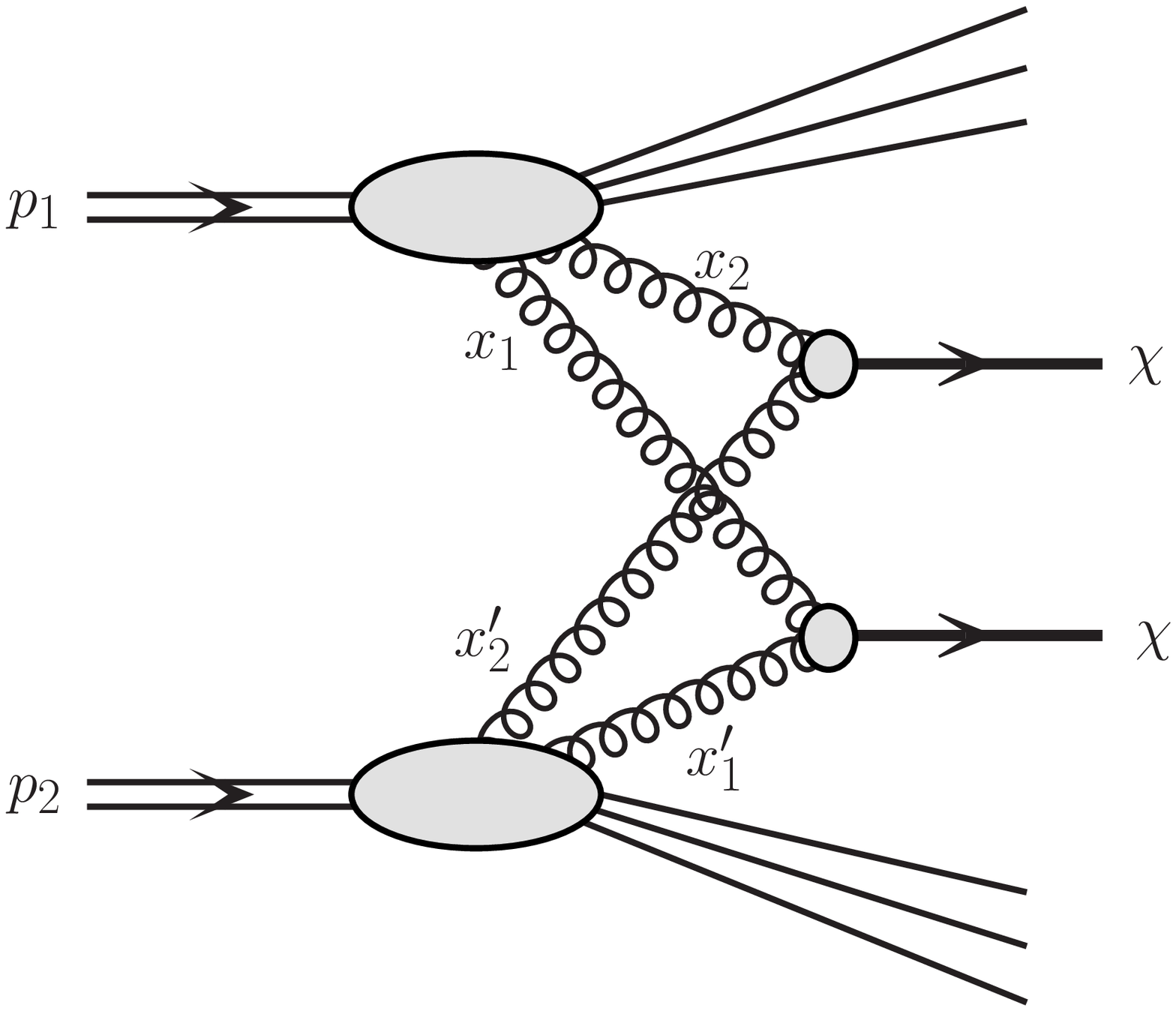}
\includegraphics[width=5cm]{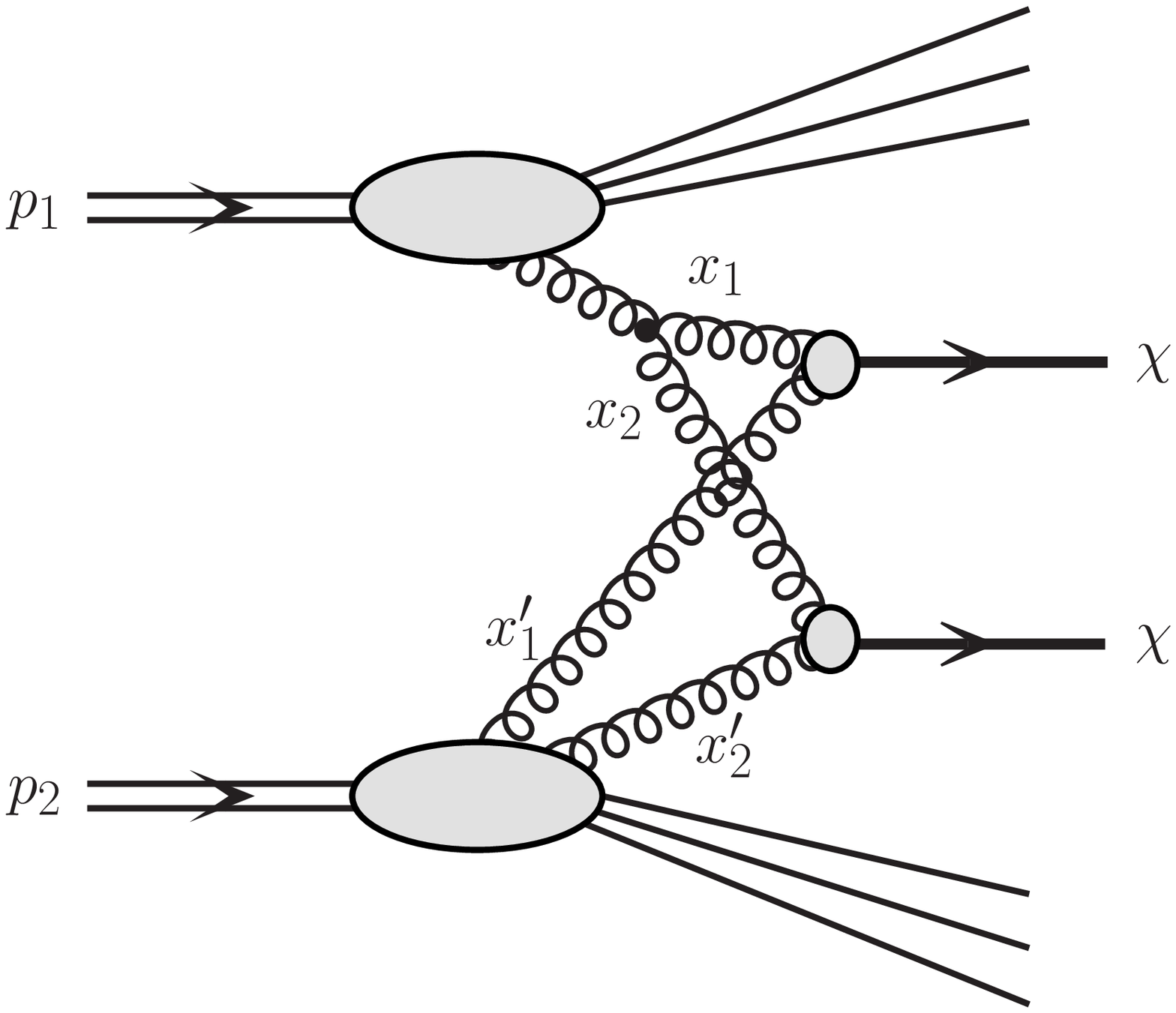}
\includegraphics[width=5cm]{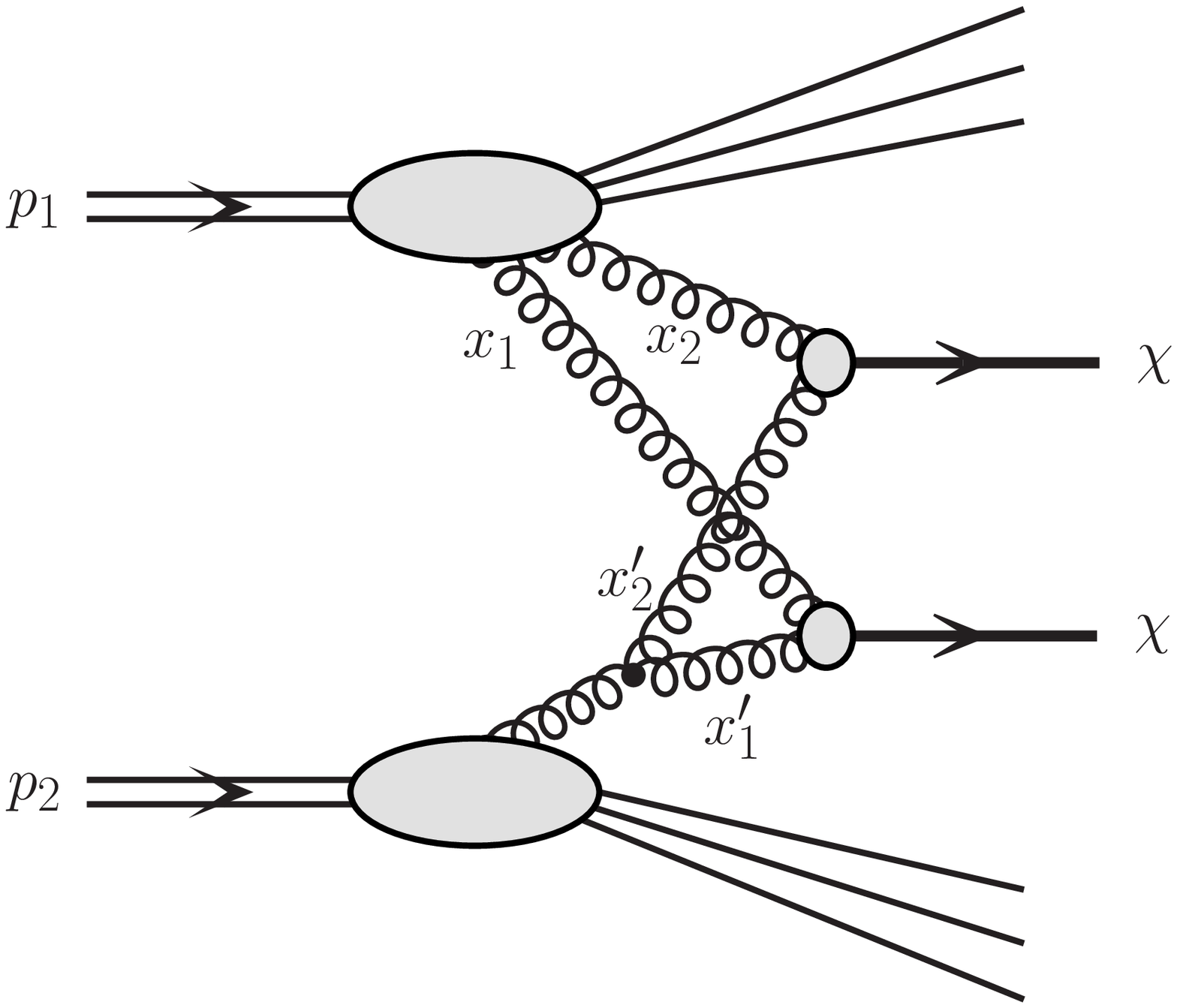}
   \caption{
\small The diagrams for DPS production of two quarkonia.
}
 \label{fig:diagrams_chichi}
\end{figure}

In Fig.~\ref{fig:diagrams_chichi} we show the 2v2 and 2v1 DPS mechanisms
of production of two quarkonia or two Higgs bosons. The first mechanism 
is the classical DPS mechanism (2v2) and the other two represent mechanisms 
(2v1) with perturbative splitting of one of the ladders.

%----------------------------------------------------------------
%----------------------------------------------------------------

As mentioned in the introduction, we ignore any possible contribution to DPS coming 
from double perturbative splitting or 1v1 graphs, and focus instead on the relative 
contributions coming from 2v1 and 2v2 graphs. Then, under certain assumptions, the 
leading-order (LO) cross section for the DPS production of two quarkonia or two Higgs 
bosons can be written in a compact way \cite{Gaunt:2012dd, BDFS2013} as
\begin{align} \label{DPSmaster}
\sigma(DPS) = \sigma(2v2) + \sigma(2v1)
\end{align}
with
\begin{eqnarray} \label{sigma2v2}
\sigma(2v2) &=& \frac{m}{2} \frac{1}{\sigma_{eff,2v2}}
\int dx_1 dx_2 dx_1' dx_2' \;
\sigma_{g g \to \chi}(x_1 x_1' s) \;
\sigma_{g g \to \chi}(x_2 x_2' s) \\ \nonumber
&& \times \; D^{gg}(x_1, x_2, \mu_1^2, \mu_2^2) D^{gg}(x_1, x_2, \mu_1^2, \mu_2^2)
\label{2v2}
\end{eqnarray}
and
\begin{eqnarray} \label{sigma2v1}
\sigma(2v1) &=& \frac{m}{2} \frac{1}{\sigma_{eff,2v1}}
\int dx_1 dx_2 dx_1' dx_2' \; \sigma_{g g \to \chi}(x_1 x_1' s) \;
\sigma_{g g \to \chi}(x_2 x_2' s) \\ \nonumber 
&& \times \; \left(
{\hat D}^{gg}(x_1', x_2', \mu_1^2, \mu_2^2) D^{gg}(x_1, x_2, \mu_1^2, \mu_2^2)
+
D^{gg}(x_1', x_2', \mu_1^2, \mu_2^2) {\hat D}^{gg}(x_1, x_2, \mu_1^2, \mu_2^2)
\right) \; ,
\label{2v1}
\end{eqnarray}
where $m$ = 1 for two identical final states and $m$ = 2 for two different
final states. The quantities $D^{ij}$ and ${\hat D}^{ij}$ are the 
independent ladder pair and ladder splitting double PDFs (dPDFs) respectively.
Roughly speaking, the first gives the probability to find a pair of partons
in the proton that was generated as a result of a pair existing at the 
nonperturbative level independently radiating partons. The second gives the
probability to find a pair of partons that was generated as a result of 
one parton perturbatively splitting into two.  We will
give more detail as to how these objects are computed shortly.

The key assumption needed to obtain \eqref{DPSmaster} is that the 2pGPD for 
the independent ladder pair can be factorised as follows:
\begin{equation} \label{2pGPDfact}
\Gamma^{ij}(x_1, x_2, \mu_1^2, \mu_2^2,b) = D^{ij}(x_1, x_2, \mu_1^2, \mu_2^2) F(b) \; ,
\end{equation}
where $F(b)$ is normalised to 1.
Since the two partons $i$ and $j$ are only connected via nonperturbative
processes we expect $F(b)$ to be some smooth function with a width 
of order of the proton radius. The quantities $\sigma_{eff,2v1}=\sigma_{eff,1v2}$ 
and $\sigma_{eff,2v2}$ in \eqref{sigma2v1}
and \eqref{sigma2v2} are related to $F(b)$ as follows:
\begin{eqnarray}
\frac{1}{\sigma_{eff,2v2}} &=& \int d^2b [F(b)]^2 \; , \\
\frac{1}{\sigma_{eff,2v1}} &=& F(b=0) \; .
\label{effective_cross_sections}
\end{eqnarray}
Under the approximation in which the independent branching partons 
are uncorrelated in transverse space, $F(b)$ is given by a convolution 
of an azimuthally symmetric transverse parton density in 
the proton $\rho(\vect{r})$ with itself, where $\rho(\vect{r})$ 
must be normalised to $1$ in order to ensure the appropriate 
normalisation of $F(\vect{b})$:
\begin{align} \label{F&rho}
F(b) = \int d^2\vect{r} \rho(\vect{r}) \rho(\vect{b-r}) \; .
\end{align}

\begin{table}
\centering
\begin{tabular}{| c | c| c |}
\hline
\, Transverse density profile \,& $\rho(r)$ &\,$\sigma_{eff,1v2}/\sigma_{eff,2v2}$\, \\
\hline
Hard Sphere  &  \, $\rho(r) = \tfrac{3}{2\pi R^2}(1-r^2/R^2)^{1/2}\Theta(R-r)$  \, &0.52 \\
Gaussian                   & $\rho(r) = \tfrac{1}{2\pi R^2}\exp\left(-\tfrac{r^2}{2R^2}\right)$  &0.50 \\
Top Hat                    &  $\rho(r) = \tfrac{1}{\pi R^2}\Theta(R-r)$  &0.46 \\
Dipole                     &  $\rho(r) = \int \tfrac{d^2\vect{\Delta}}{(2\pi)^2}e^{i\vect{\Delta} \cdot \vect{r}}(\Delta^2/m_g^2+1)^{-2}$  &0.43 \\
Exponential                &  $\rho(r) = \int dz \tfrac{1}{8\pi R^3} \exp(-\sqrt{r^2+z^2}/R)$   &0.43\\
\hline
\end{tabular}
\caption{\label{tab:SigEffR} Ratio of $\sigma_{eff,1v2}$ to $\sigma_{eff,2v2}$ for various simple choices for the proton
transverse density profile $\rho(r)$, under the approximations \eqref{2pGPDfact} and \eqref{F&rho} introduced in
the text. The hard sphere projection and exponential profiles are studied in \mycite{Domdey:2009bg}, and the dipole
profile is studied in \mycites{Frankfurt:2003td, Blok:2010ge, RS2011, Blok:2011bu}. $R$ and $m_g$ are constants which do
not affect the $\sigma_{eff,1v2}/\sigma_{eff,2v2}$ ratio. 
}
\end{table}

In a simple model where $\rho(\vect{r})$ is taken to have Gaussian
functional form one gets $\sigma_{eff,2v1} = \sigma_{eff,2v2}/2$. 
Other simple functional forms for $\rho(\vect{r})$ also with one width 
parameter yield similar results, as illustrated in Table
\ref{tab:SigEffR}. Using a model with two width parameters for 
$\rho(\vect{r})$, one obtains an enhancement of the ratio 
$\sigma_{eff,2v2}/\sigma_{eff,2v1}$ as one of the width 
parameters becomes small compared to the other, and the distribution 
becomes `clumpy', although this enhancement is rather weak unless one 
chooses an extremely clumpy distribution.
In order to illustrate this, we use the `triple hot spot' model
described in section 4 of \mycite{Domdey:2009bg} for the independent 
ladder pair transverse density (see also 
\mycites{Bender:1983cw, Calucci:1999yz, Povh:2002vg}). In this model, 
the proton contains three clumps of parton density which can be thought 
of as the three gluon clouds surrounding the valence quarks, and 
$F(\vect{b})$ given by:

\begin{align}
F(\vect{b}) =& \frac{1}{4}\int d^2{\vect b}_{1}
d^2{\vect
b}_{v_1}\, d^2{\vect b}_{v_2} \vert\psi\left({\vect b}_{v_1},{\vect
b}_{v_2}\right)\vert^2\, \sum_{ij}^2 d({\vect b}_1,{\vect
b}_{v_i})\, d({\vect b}_1-\vect{b},{\vect b}_{v_j}). \label{eq4.3}
\end{align}
where:
\begin{align}
 |\psi({\vect b}_{v_1},{\vect b}_{v_2})|^2
=&\frac{3}{\pi^2\delta_v^4}
\exp\left[-\frac{1}{3\delta_v^2}\left(({\vect b}_{v_1}-{\vect
b}_{v_2})^2+({\vect b}_{v_1}-{\vect b}_{v_3})^2+({\vect b}_{v_2}-{\vect
b}_{v_3})^2\right)\right] 
\Bigg\vert_{ -{\vect b}_{v_3} \equiv {\vect b}_{v_1} + {\vect b}_{v_2}}
\label{eq4.1}
\\
d({\vect b},{\vect b}_v)=&\frac{1}{2\pi\delta_s^2}
\exp\left(-\frac{({\vect b}_{v}-{\vect b})^2}{2\delta_s^2}\right)\, .
\label{eq4.2}
\end{align}

There are two parameters $\delta_v$ and $\delta_s$ in the model, the first of which determines the spacing between
the hot spots, and the second of which determines the width of the hot spots. In \mycite{Domdey:2009bg}, $\delta_s$
is taken to be a function of momentum fraction $x$, but we will simply take it to be a constant here.
We can readily obtain an analytic expression for $\sigma_{eff,1v2}/\sigma_{eff,2v2}$ in this model, which only 
depends on the ratio $\delta^2_s/\delta^2_v$:
\begin{equation} \label{tricoreR}
\dfrac{\sigma_{eff,1v2}}{\sigma_{eff,2v2}} = {\frac {32
    {\delta_s^4/\delta_v^4}+16 \delta_s^2/\delta_v^2+1}{ 4\left( 4
      \delta_s^2/\delta_v^2+1 \right) ^{2}}}
\; .
\end{equation}

\begin{figure}
\centering
\includegraphics[trim = 2cm 0 0 0]{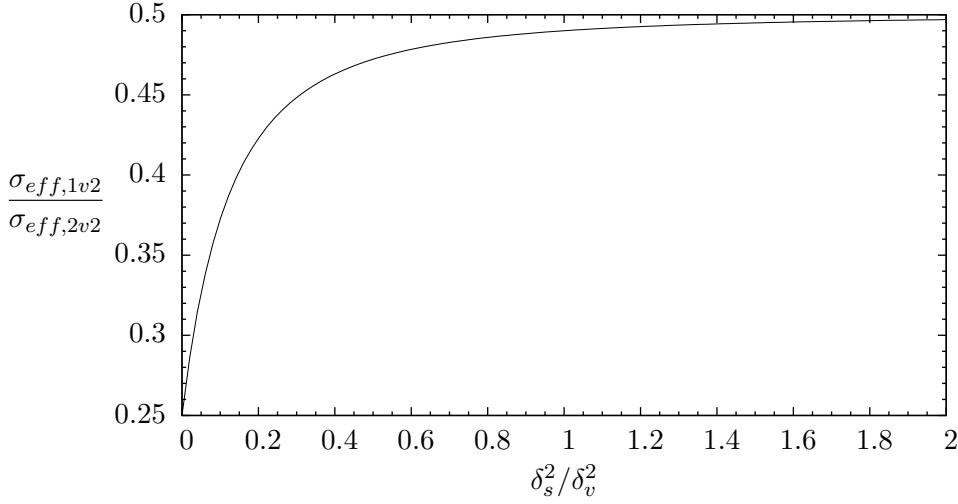}
\caption{\label{fig:tricore} Dependence of $\sigma_{eff,1v2}/\sigma_{eff,2v2}$ on $\delta_s^2/\delta_v^2$ in the 
triple hot spot model described in \mycite{Domdey:2009bg}. We have taken $\delta_s$ and $\delta_v$ not to depend on
longitudinal momentum fractions $x_i$.
}
\end{figure}

This function is plotted in Fig.~\ref{fig:tricore} for $\delta^2_s/\delta^2_v$ values between $0$ and $2$. The
function value never exceeds 0.5, and asymptotes to the single Gaussian result of 0.5 as $\delta_s$ becomes very
much larger than $\delta_v$. As $\delta^2_s/\delta^2_v$ is reduced (corresponding to the lumps in the transverse
density becoming more pronounced), $\sigma_{eff,1v2}/\sigma_{eff,2v2}$ decreases as anticipated, reaching $0.25$ at 
$\delta^2_s/\delta^2_v=0$. In practice taking $\delta^2_s/\delta^2_v$ smaller than perhaps $\sim 0.1$ is not reasonable 
(given that it is supposed to correspond to the area of a nonperturbative gluon lump divided by the area of a proton), 
and imposing this constraint we find $0.372 < \sigma_{eff,1v2}/\sigma_{eff,2v2} < 0.5$.

Therefore we see that there is a geometrical enhancement of the 2v1 contributions
with respect to the 2v2 ones, and if the approximations \eqref{2pGPDfact} and \eqref{F&rho} are valid, then this enhancement should
be rather close to a factor of 2, as was first emphasised in \mycite{Blok:2011bu}. % We shall discuss this in more detail in the Result section.

The independent ladder pair and ladder splitting dPDFs, 
$D^{ij}(x_1, x_2, \mu_1^2, \mu_2^2)$ and ${\hat D}^{ij}(x_1, x_2, \mu_1^2, \mu_2^2)$,
are calculated as follows.

Let us begin with the ladder splitting double PDF, and consider
the case in which the scales are equal: ${\hat D}^{ij}(x_1, x_2, \mu^2, \mu^2) 
\equiv {\hat D}^{ij}(x_1, x_2, \mu^2)$. This is initiated at zero at
some low scale $Q_0$:
\begin{align} \label{1t2bcs}
{\hat D}^{{j_1}{j_2}}(x_1,x_2;\mu^2 = Q_0^2) = 0 \; .
\end{align}
$Q_0$ is the scale at which perturbative $1 \to 2$ splittings begin to
occur, which should be of order of $\Lambda_{QCD}$. The ladder splitting
dPDF ${\hat D}^{ij}(x_1, x_2, \mu^2)$ evolves according to the double DGLAP
equation of \mycites{SSZ82, GS2010}:
\begin{align} \label{dbDGLAP}
\mu^2 \dfrac{d{\hat D}^{{j_1}{j_2}}(x_1,x_2;\mu^2)}{d\mu^2} = \dfrac{\alpha_s(\mu^2)}{2\pi} \Biggl[
\sum_{j'_1}\int_{x_1}^{1-x_2}\dfrac{dx'_1}{x'_1}{\hat D}^{{j'_1}{j_2}}(x'_1,x_2;\mu^2)P_{j'_1
\to j_1}\left(\dfrac{x_1}{x'_1}\right)
\nonumber\\
+\sum_{j'_2}\int_{x_2}^{1-x_1}\dfrac{dx'_2}{x'_2}{\hat D}^{{j_1}{j'_2}}(x_1,x'_2;\mu^2)P_{j'_2
\to j_2}\left(\dfrac{x_2}{x'_2}\right)
\nonumber\\
+\sum_{j'}D^{j'}(x_1+x_2;\mu^2)\dfrac{1}{x_1+x_2}P_{j' \to j_1
j_2}\left(\dfrac{x_1}{x_1+x_2}\right) \Biggr] \; .
\end{align}

In order to calculate the ladder splitting for $\mu_1^2 > \mu_2^2$ (say),
we start from the equal scale case and then evolve up in $\mu_1^2$ using
the following evolution equation:
\begin{align} \label{1t2uneq}
\mu_1^2 \dfrac{d{\hat D}^{{j_1}{j_2}}(x_1,x_2;\mu_1^2,\mu_2^2)}{d\mu_1^2} = \dfrac{\alpha_s(\mu_1^2)}{2\pi} \Biggl[
\sum_{j'_1}\int_{x_1}^{1-x_2}\dfrac{dx'_1}{x'_1}{\hat D}^{{j'_1}{j_2}}(x'_1,x_2;\mu_1^2,\mu_2^2)P_{j'_1
\to j_1}\left(\dfrac{x_1}{x'_1}\right) \Biggr]
\end{align}
which only applies when $\mu_1^2 > \mu_2^2$. This equation is the equivalent of
equation (9) in \mycite{Ceccopieri:2010kg} (except there the evolution with respect
to $\mu_2^2$ is presented when $\mu_2^2 > \mu_1^2$, so that equation differs from
\eqref{1t2uneq} by swapping the $1$ and $2$ indices). It is straightforward to show
that equations \eqref{1t2bcs}, \eqref{dbDGLAP} and \eqref{1t2uneq} are equivalent
to equation (2.33) in \mycite{Gaunt:2012dd}.

To solve the differential equations \eqref{dbDGLAP} and \eqref{1t2uneq} and 
obtain the ladder splitting dPDFs in practice we use the numerical code of 
\mycite{GS2010}. A grid of dPDF values covering the ranges $1 \text{ GeV}^2 < 
\mu_1^2,\mu_2^2 < 500^2 \text{ GeV}^2$, $10^{-6} < x_1,x_2 < 1$ was generated 
using 300 points in the $x$ direction, and 60 points in the $\log(\mu^2)$ direction
for the evolution. We use the MSTW 2008 LO single PDFs \cite{Martin:2009iq} 
as the single PDFs in the evolution. Since the starting scale for these PDFs 
is $1$ GeV, we are not able to take $Q_0$ lower than this value, and in fact 
we set $Q_0 = 1$ GeV. We use the same $\alpha_s$ and variable flavour number 
scheme as MSTW 2008 LO, with $m_c = 1.40$ GeV and $m_b = 4.75$ GeV.

For the independent pair distribution $D^{ij}(x_1, x_2, \mu_1^2, \mu_2^2)$ 
we must specify some nonperturbative input distributions at the input scale
$\mu_1^2 = \mu_2^2 = Q_0^2$. Normally, due to the lack of information about
the dPDFs, one commonly takes the input distributions to be a product of 
single PDFs:
\begin{equation} \label{2t2input}
 D^{ij}(x_1, x_2, \mu_1^2=Q_0^2, \mu_2^2=Q_0^2) = 
 D^{i}(x_1, Q_0^2) D^{j}(x_2, Q_0^2) \; .
\end{equation}

Strictly speaking, this input should then be evolved up in scale using 
\eqref{dbDGLAP} with the final inhomogeneous term removed, and then 
\eqref{1t2uneq} when $\mu_1^2 > \mu_2^2$. However, this evolution is
almost equivalent to independent DGLAP evolution of the two partons, up
to effects of the kinematic limit in the homogeneous double DGLAP 
evolution (which manifest themselves in equations \eqref{dbDGLAP} and
\eqref{1t2uneq} by the limits of the $x'$ integrations being $1-x_i$
rather than $1$). This kinematic effect is known to be small unless 
$x$ is rather large \cite{DKK2014}, so if we take \eqref{2t2input}
as our input distributions, then to a good approximation, we can say:
\begin{equation} \label{2t2ev}
 D^{ij}(x_1, x_2, \mu_1^2, \mu_2^2) \simeq 
D^{i}(x_1, \mu_1^2) D^{j}(x_2, \mu_2^2) \; .
\end{equation}

Here we use \eqref{2t2ev} for the independent pair dPDFs. We take
the single PDFs in this equation to be the MSTW 2008 LO PDFs for consistency 
with the ladder splitting dPDFs.

We should point out that in our study we ignore several effects. The first
of these is crosstalk between the nonperturbatively generated ladder pair 
in the 2v1 graphs. This was first noticed in \mycite{Gaunt:2012dd} but was also
shown there to be numerically small in practice, so we can safely ignore it.
We also ignore effects associated with correlations or interference in spin, 
colour, flavour, fermion number, and parton type between the two partons
\cite{Mekhfi:1985dv, DOS2012}. Colour, fermion number and parton type 
correlations/interference are known to be Sudakov suppressed \cite{Mekhfi:1988kj,
DOS2012, Manohar:2012jr}, but could potentially be non-negligible for small scales
of order of a few GeV (see figure 10 of \mycite{Manohar:2012jr}). Spin correlations 
were studied in \mycite{DKK2014} in the context of the 2v2 process, and were found
to be rather small after evolution, especially when both partons in $D^{ij}$ were
gluons. They were reduced to a few tens of per cent of the unpolarised contribution after 
only a few GeV of evolution, even in the most optimistic input scenario.
However, it might be interesting to do a more detailed study of the spin effects, 
also including their effect in the 2v1 graphs. This is particularly in light of the
experimentally observed azimuthal correlations between two $D^0$ mesons produced in
proton-proton collisions \cite{LHCb}, which cannot be described using an unpolarised DPS mechanism
(either 2v2 or 2v1) \cite{MS2013,HMS2014}.
For the gluon-initiated processes we will discuss here, quark flavour interference
is not a relevant effect since the flavour interference distributions are not able
to mix with the double gluon distribution. Finally, we ignore interference between
DPS and SPS, or twist 3 vs twist 3 terms, which were discussed in \mycites{DOS2012, 
Manohar:2012jr}. It is possible to show that some of the twist 3 vs twist 3
effects are suppressed by helicity nonconservation in the associated diagrams
\cite{Manohar:2012jr, Qiu:1990xxa}, but it seems likely that not all such effects 
are suppressed in this way -- this topic needs further study.

In this paper we discuss production processes for which gluon-gluon fusion
is the dominant process. We begin with processes $gg \to A$ in which a single
particle $A$ is produced from the hard scattering process ($A = H, \eta, \chi, ...$).
Then, at the leading order to which we work in this paper:
\begin{equation}
\sigma_{gg \to \chi}(\hat{s}) = C_{g g \to \chi} \cdot 
\delta({\hat s} - M_{\chi}^2)
\; .
\end{equation}
This allows us to simplify considerably the cross section.
In this approximation one can easily get the cross section 
differential in rapidity of one and second object (meson or Higgs boson).

\begin{eqnarray}
\sigma(2v2) &=& \frac{m}{2} \frac{1}{\sigma_{eff,2v2}}
\int dy_1 dy_2 \; C_{g g \to \chi}^2 \; x_1 x_1' x_2 x_2' %\\ \nonumber
%&& \times \;
D^{gg}(x_1, x_2, \mu_1^2, \mu_2^2) \; \nonumber
D^{gg}(x_1, x_2, \mu_1^2, \mu_2^2)\\ 
\label{simplified_2v2}         
\end{eqnarray}
and
\begin{eqnarray}
\sigma(2v1) &=& \frac{m}{2} \frac{1}{\sigma_{eff,2v1}}
\int dy_1 dy_2 \; C_{g g \to \chi}^2 \; x_1 x_1' x_2 x_2' \\ \nonumber
&& \times \;
\left(
{\hat D}^{gg}(x_1', x_2', \mu_1^2, \mu_2^2) D^{gg}(x_1, x_2, \mu_1^2, \mu_2^2)
+
D^{gg}(x_1', x_2', \mu_1^2, \mu_2^2) {\hat D}^{gg}(x_1, x_2, \mu_1^2, \mu_2^2)
\right) \; .
\label{simplified_2v1}
\end{eqnarray}
This allows to easily calculate distributions in rapidity
of $\chi_A$ and $\chi_B$.
In the last two equations the longitudinal momentum fractions
are calculated from masses of the produced objects 
(quarkonia, Higgs bosons) and their rapidities
\begin{eqnarray}
x_1  &=& \frac{M}{\sqrt{s}} \exp( y_1 ), \;\;\;
x_1'  =  \frac{M}{\sqrt{s}} \exp(-y_1 ), \nonumber \\
x_2  &=& \frac{M}{\sqrt{s}} \exp( y_2 ), \;\;\;
x_2'  =  \frac{M}{\sqrt{s}} \exp(-y_2 ).
\label{x_definition}
\end{eqnarray}
%

%------------------------------------------------------------
\subsection{DPS production of $\bm{c \bar c c \bar c}$} \label{sec:DPScccc}
%------------------------------------------------------------

\begin{figure}[!h]
\includegraphics[width=5cm]{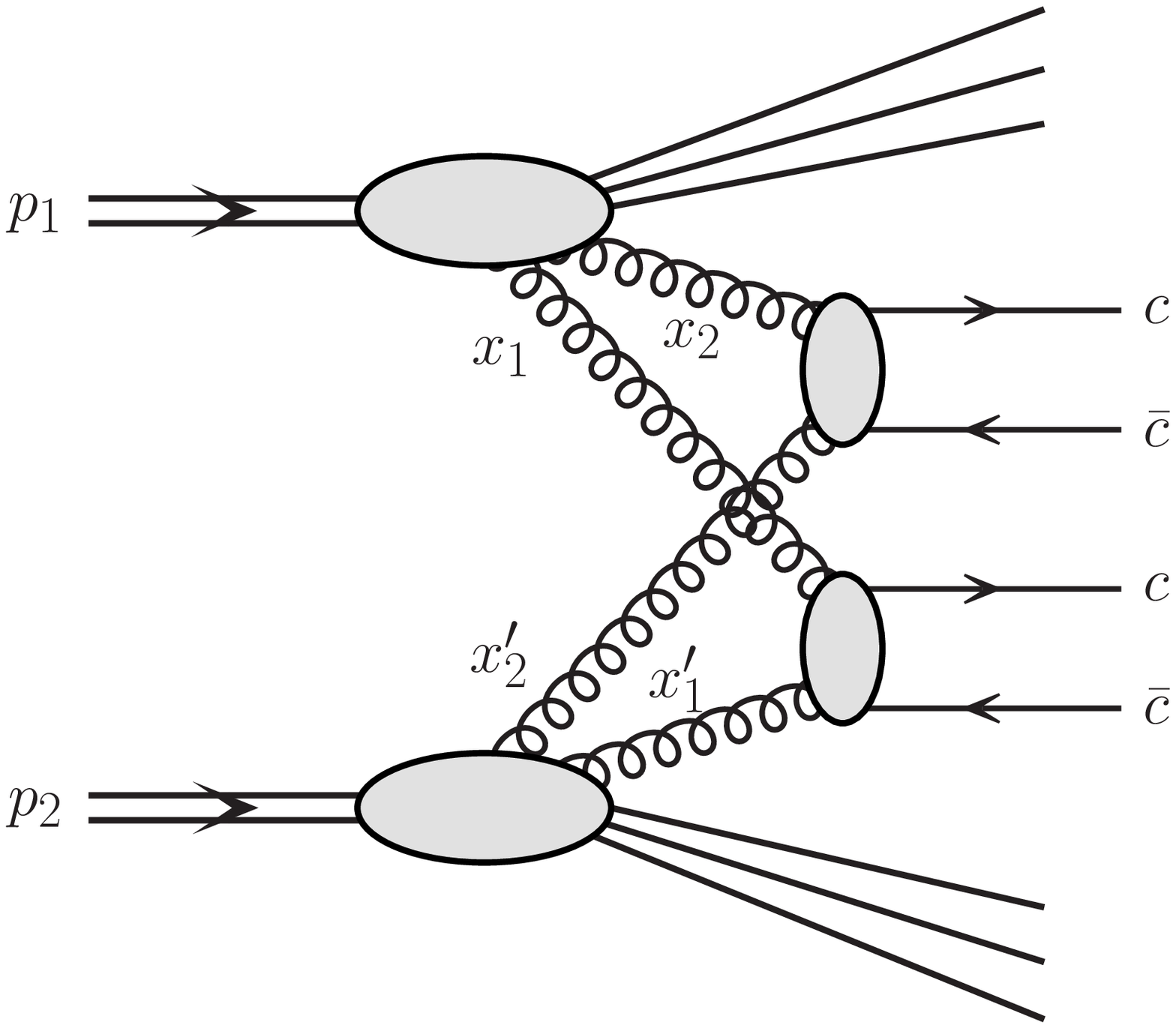}
\includegraphics[width=5cm]{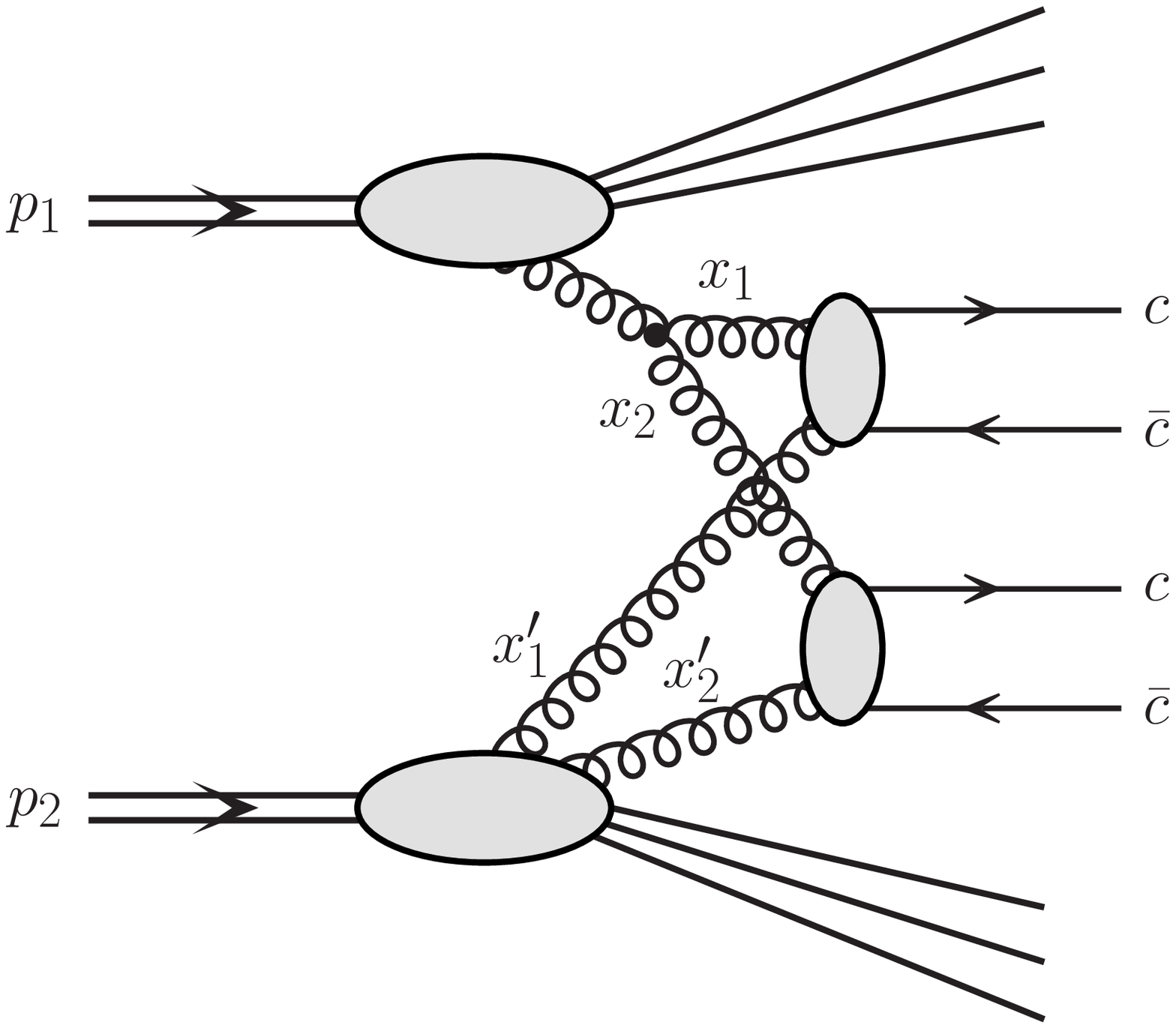}
\includegraphics[width=5cm]{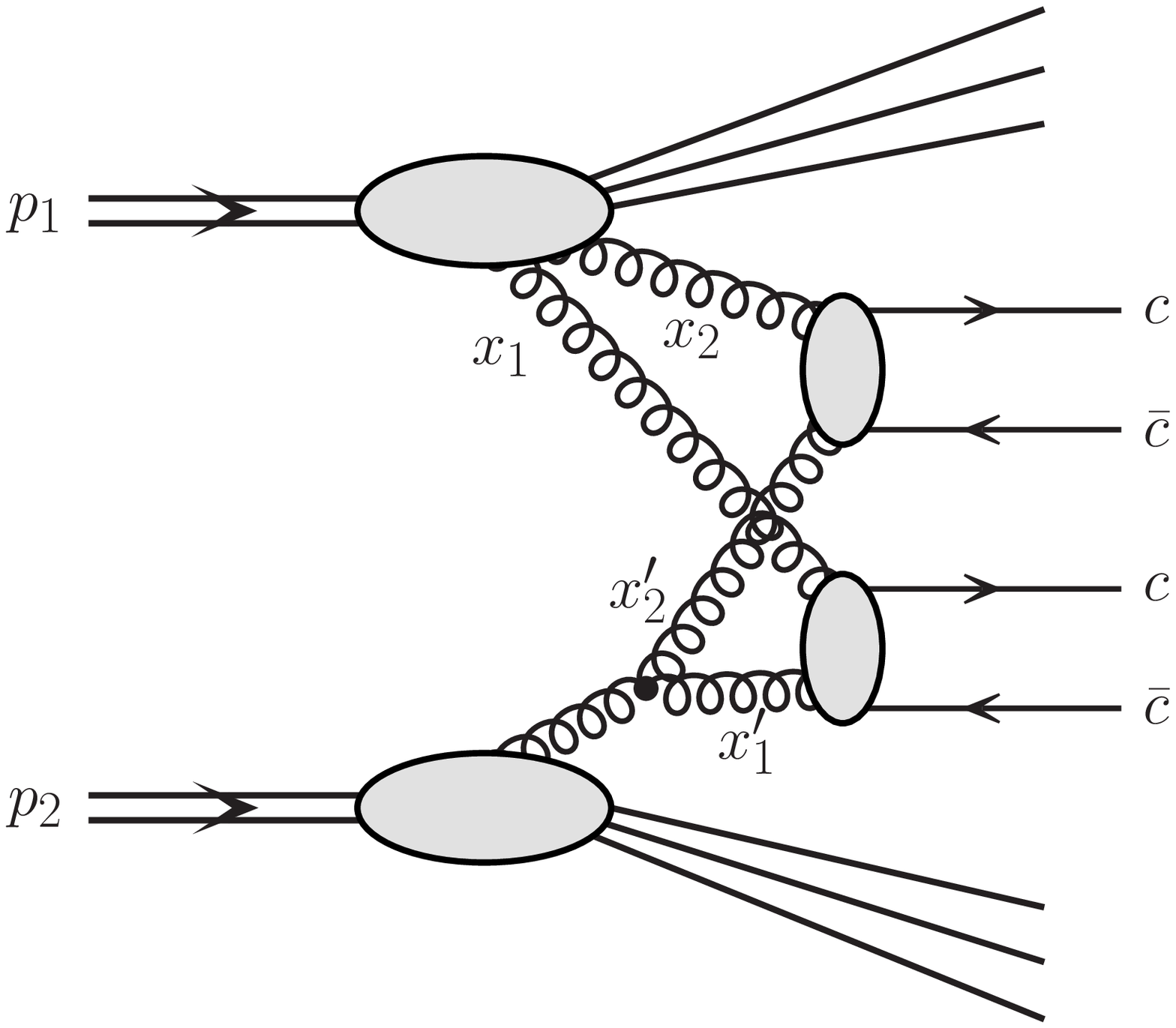}
   \caption{
\small The diagrams for DPS production of $c \bar c c \bar c$.
}
 \label{fig:diagrams_ccbarccbar}
\end{figure}

In Fig.~\ref{fig:diagrams_ccbarccbar} we show similar DPS mechanisms
for $c \bar c c \bar c$ production. The 2v1 mechanism 
(the second and third diagrams) were not considered so far 
in the literature.

%----------------------------------------------------------------
%----------------------------------------------------------------

In contrast to double quarkonium production in the case of 
$c \bar c c \bar c$ production the cross section formula is a bit 
more complicated and the kinematical variables of each produced
particle ($c$ quark or $\bar c$ antiquark) must be taken 
into account
\begin{eqnarray}
\sigma(2v2) &=& \frac{1}{2} \frac{1}{\sigma_{eff,2v2}}
\int dy_1 dy_2 d^2 p_{1t} dy_3 dy_4 d^2 p_{2t} 
\frac{1}{16 \pi {\hat s}^2} \overline{ |{\cal M}(gg \to c \bar c)|^2} 
\; x_1 x_1' x_2 x_2' \\ \nonumber
&& \times \;
D^{gg}(x_1, x_2, \mu_1^2, \mu_2^2) \;
D^{gg}(x_1, x_2, \mu_1^2, \mu_2^2) 
\label{DPS_ccbarccbar_2v2}         
\end{eqnarray}
and
\begin{eqnarray} 
\sigma(2v1) &=& \frac{1}{2} \frac{1}{\sigma_{eff,2v1}}
\int dy_1 dy_2 d^2 p_{1t} dy_3 dy_4 d^2 p_{2t}  
\frac{1}{16 \pi {\hat s}^2} \overline{ |{\cal M}(gg \to c \bar c)|^2} 
\; x_1 x_1' x_2 x_2' \\ \nonumber
&& \times \;
\left(
{\hat D}^{gg}(x_1', x_2', \mu_1^2, \mu_2^2) D^{gg}(x_1, x_2, \mu_1^2, \mu_2^2)
+
D^{gg}(x_1', x_2', \mu_1^2, \mu_2^2) {\hat D}^{gg}(x_1, x_2, \mu_1^2, \mu_2^2)
\right) 
\label{DPS_ccbarccbar_2v1}
\end{eqnarray}
for conventional and perturbative splitting contributions, respectively.
The integration is 6-fold. The same is true for differential distributions.
In the last two equations the longitudinal momentum fractions
are calculated from the transverse masses $m_{t}$ of the produced 
quarks/antiquarks and their rapidities
\begin{eqnarray} \label{ccccxs}
x_1  &=& \frac{m_{1t}}{\sqrt{s}} (\exp( y_1 ) + \exp( y_2 )), \;
x_1'  =  \frac{m_{1t}}{\sqrt{s}} (\exp(-y_1 ) + \exp(-y_2 )), \nonumber \\
x_2  &=& \frac{m_{2t}}{\sqrt{s}} (\exp( y_3 ) + \exp( y_4 )), \;
x_2'  =  \frac{m_{2t}}{\sqrt{s}} (\exp(-y_3 ) + \exp(-y_4 )).
\label{x_definition_ccbarccbar}
\end{eqnarray}

The quantity $m_{1t}$ corresponds to the transverse mass of either parton
produced from the first hard subprocess, whilst $m_{2t}$ corresponds to that from
the second hard subprocess. The transverse mass $m_t$ is defined in the usual way
to be $\sqrt{p_\perp^2+m^2}$.

%------------------------------------------------------------------------------
\subsection{Energy and process dependence 
of the effective cross section 
due to the presence of the perturbative splitting}
%------------------------------------------------------------------------------

\label{subsection:energy_dependence}

The cross section for DPS production of some final states
(e.g. $\chi$,$\chi$ or $c \bar c, c \bar c$)
can be written in a somewhat simplified way as:
\begin{equation}
\sigma^{DPS} = \frac{1}{\sigma_{eff,2v2}} \Omega^{2v2}
             + \frac{1}{\sigma_{eff,2v1}} \Omega^{2v1}  \; .
\end{equation}
$\Omega^{2v2}$ and $\Omega^{2v1}$ contain the $D$ functions and
cross section of a process chosen\footnote{In general above $\sigma_{eff,2v2}$, $\sigma_{eff,2v1}$
and $\sigma^{DPS}$ can be differential in $x$'s as well as
can represent partially or fully phase space integrated quantities.}.
The equation is true both for phase space integrated cross section
and differential distributions.
The equation reflects the presence of the two components (2v2 and 2v1)
as discussed above.

In phenomenology this is often simplified and written as
\begin{equation}
\sigma^{DPS} = \frac{1}{\sigma_{eff}} \Omega^{2v2} \; .
\end{equation}
From the two equations above one gets:
\begin{equation}
\frac{1}{\sigma_{eff}} = \frac{1}{\sigma_{eff,2v2}} 
    + \frac{1}{\sigma_{eff,2v1}} \frac{\Omega^{2v1}}{\Omega^{2v2}} \; .
\label{sigma_eff}
\end{equation}
If we assume that in addition $\sigma_{eff,2v1} = \sigma_{eff,2v2}/2$
one gets:
\begin{equation}
\frac{1}{\sigma_{eff}} = \frac{1}{\sigma_{eff,2v2}}
\left( 1 + 2 \Omega^{2v1}/ \Omega^{2v2} \right) \; .
\label{sigma_eff_simplified_formula}
\end{equation}
As will be discussed in this paper the ratio
$\Omega^{2v1} / \Omega^{2v2}$ depends on the centre-of-mass energy
and process considered.
This means that $\sigma_{eff}$ as found from phenomenological analyses
of the data (see \mycites{Seymour2013,Bahr2013}) may depend on the energy 
as well as process considered. We shall discuss this in the Result section.

In early phenomenological estimates of $\sigma_{eff}$ that took into 
account only the 2v2 mechanism \cite{Frankfurt:2003td, Calucci:1999yz},
values of the order 30 mb were found. This is twice as large as the typical 
$\sigma_{eff}$ values found in the experimental studies ($\sigma_{eff} \sim$ 15 mb) \cite{Akesson:1986iv,Abe:1997bp,Abe:1997xk, Abazov:2009gc, Aad:2013bjm,Chatrchyan:2013xxa,Aad:2014rua}.
In \mycite{BDFS2013} it was argued that this discrepancy can be explained by
the 2v1 mechanism. We also find a similar enhancement of the DPS cross section
(i.e. reduction of $\sigma_{eff}$) by a factor of 2 coming from the 2v1 mechanism, 
as discussed below.

% Some simple phenomenological estimates of $\sigma_{eff}$ were done
% e.g. in Ref.~\cite{Gustaffson,BDFS2013}. In gluon initiated processes
% %
% \begin{equation}
% \frac{1}{\sigma_{eff}} = \int F_{2g}^2(q) \frac{d^2 q}{(2 \pi)^2} \; .
% \end{equation}
% %
% $F_{2g}(q)$ can be estimated from e.g. photoproduction of $J/\psi$
% at HERA.

%--------------------
\section{Results}
%--------------------

%-----------------------------------------------------
\subsection{Double parton distributions}
%-----------------------------------------------------

\begin{figure}[!h]
\includegraphics[width=7.3cm]{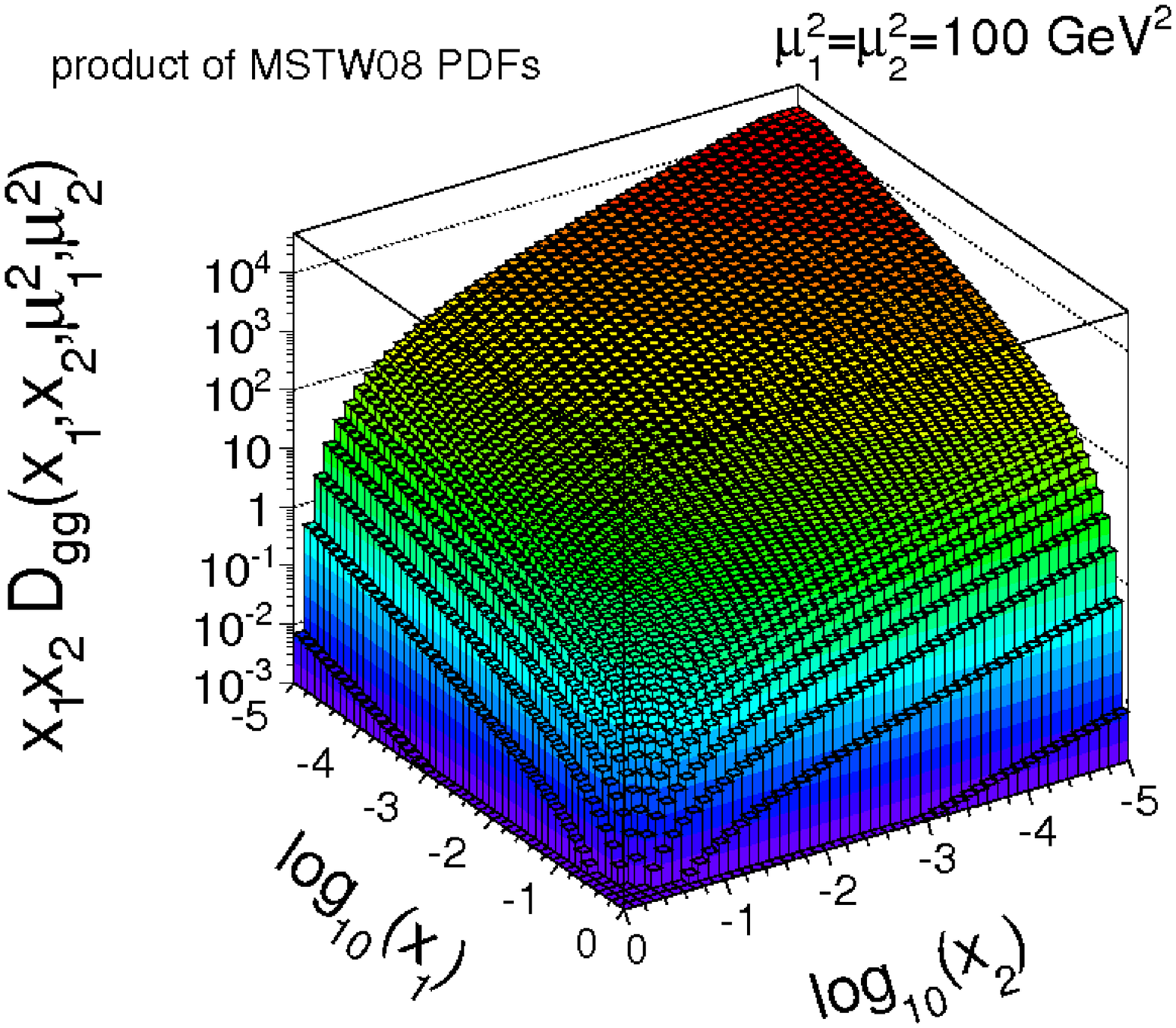}
\includegraphics[width=7.3cm]{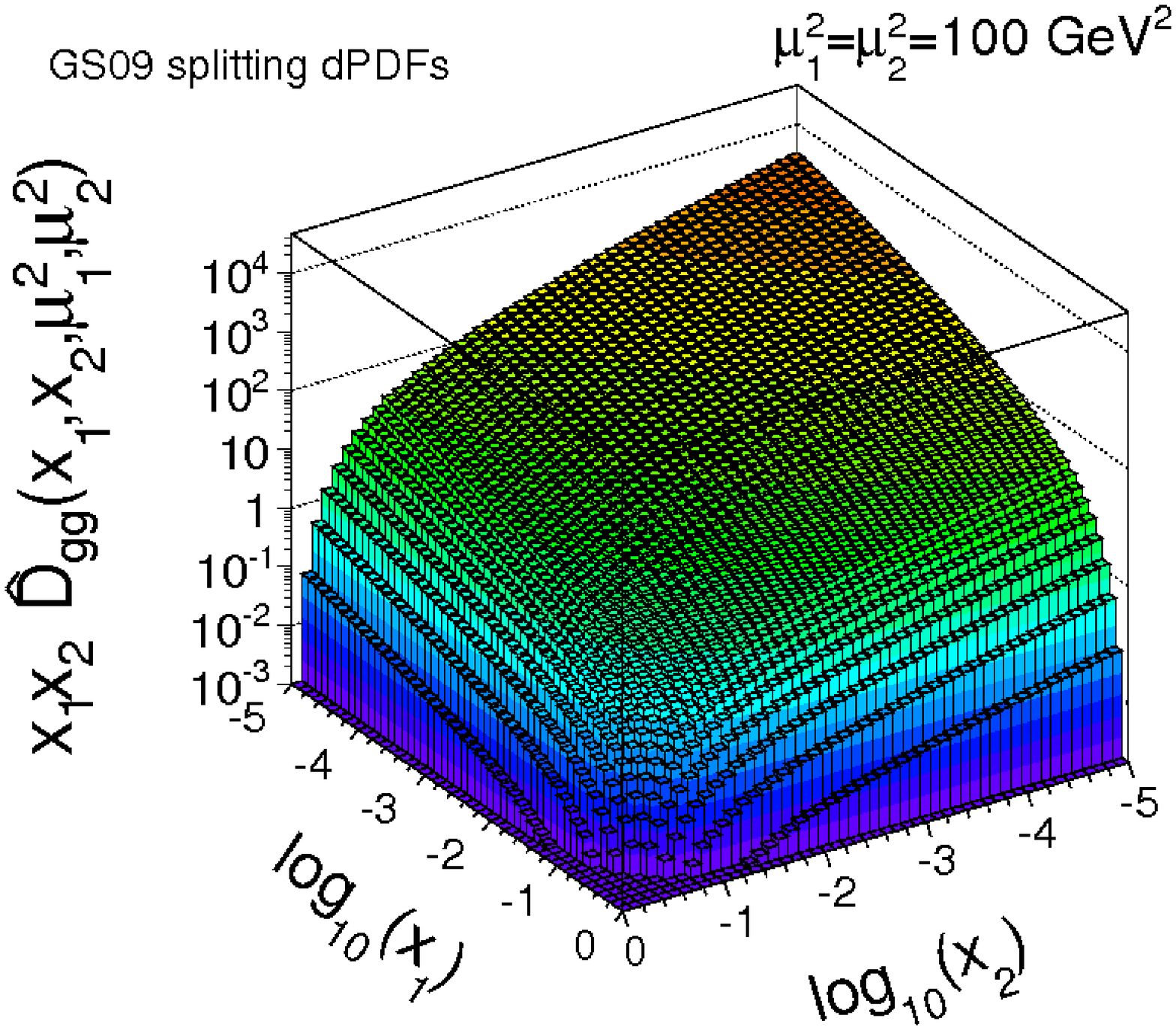} \\
\includegraphics[width=7.3cm]{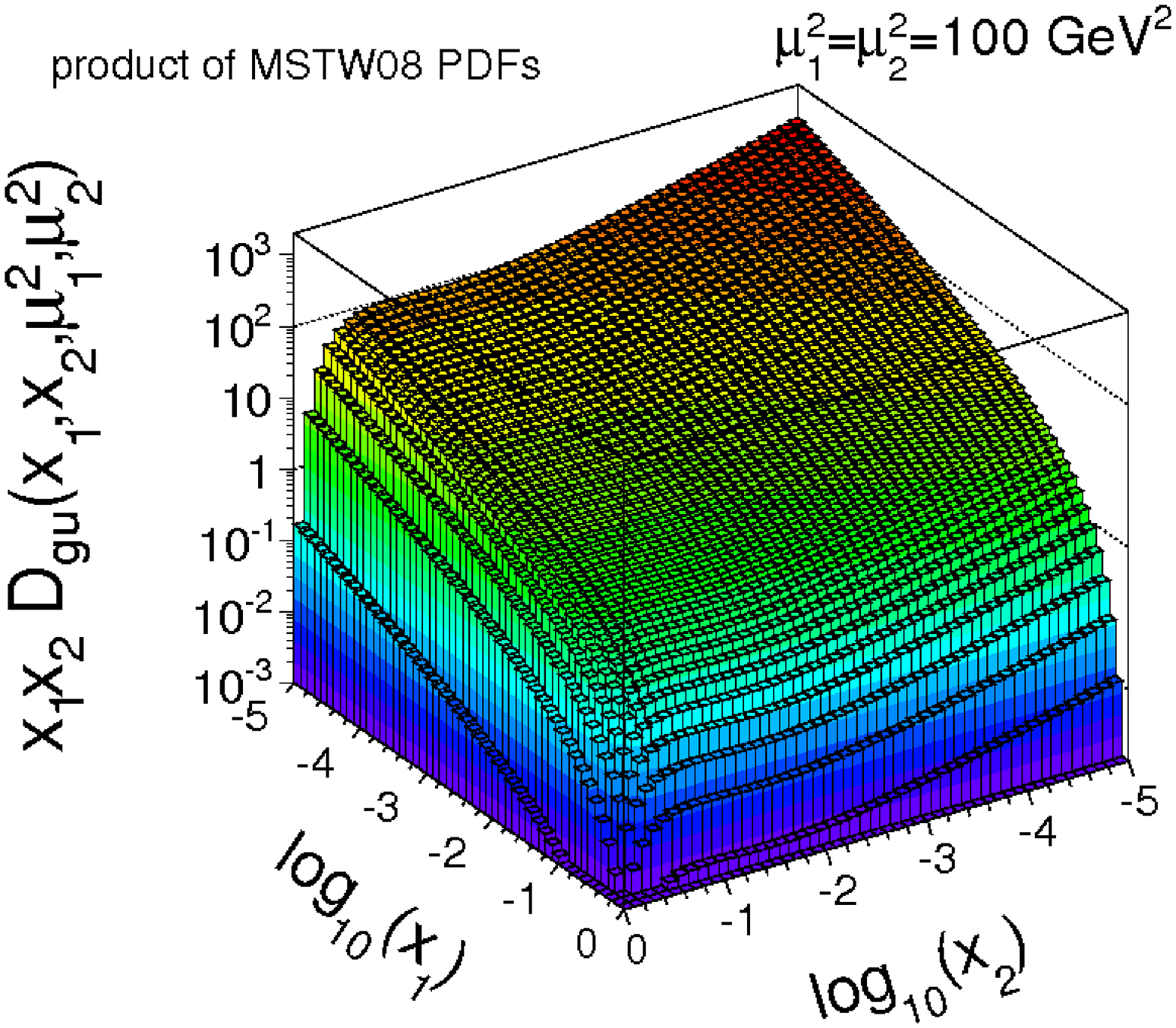}
\includegraphics[width=7.3cm]{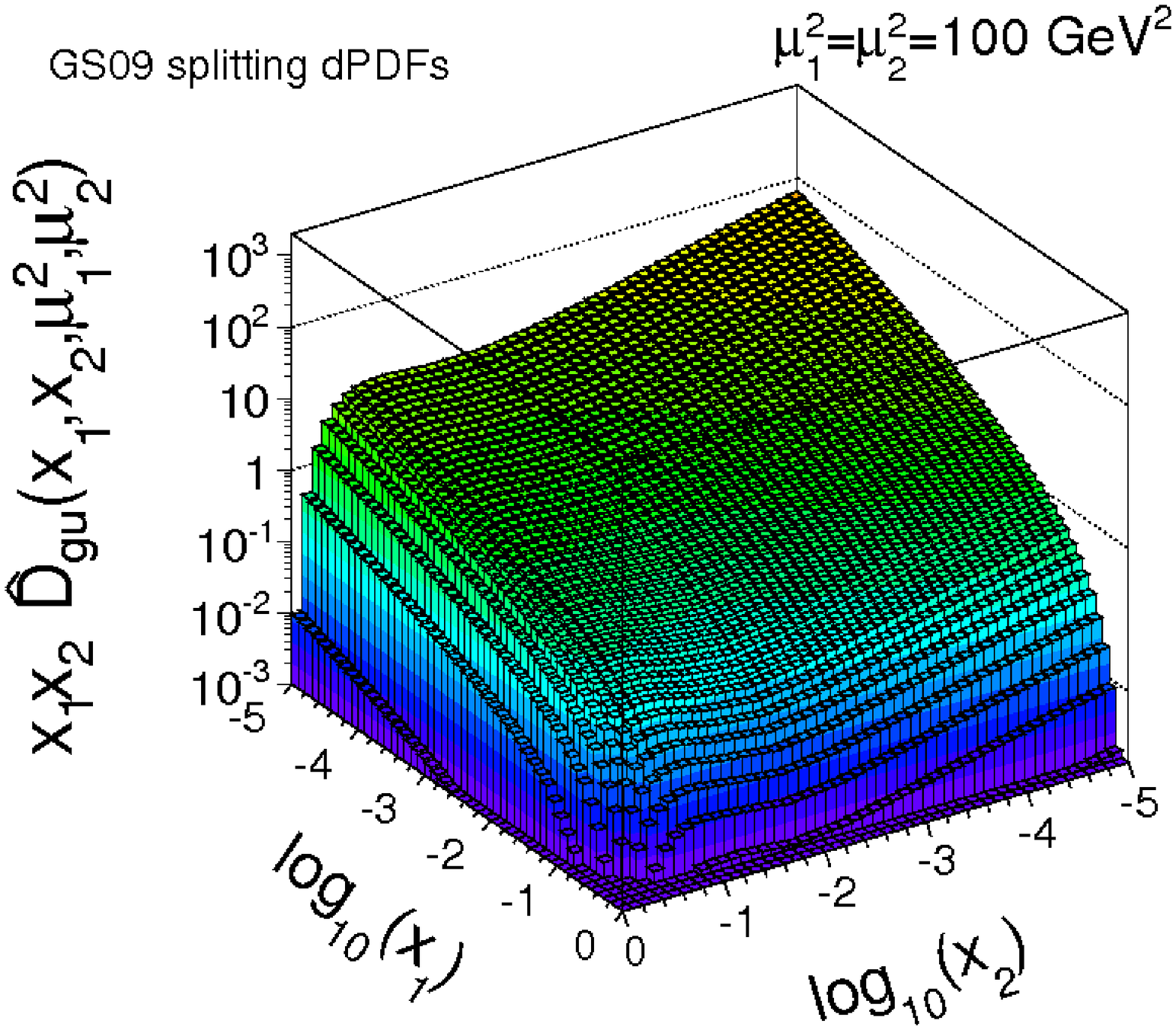} \\
\includegraphics[width=7.3cm]{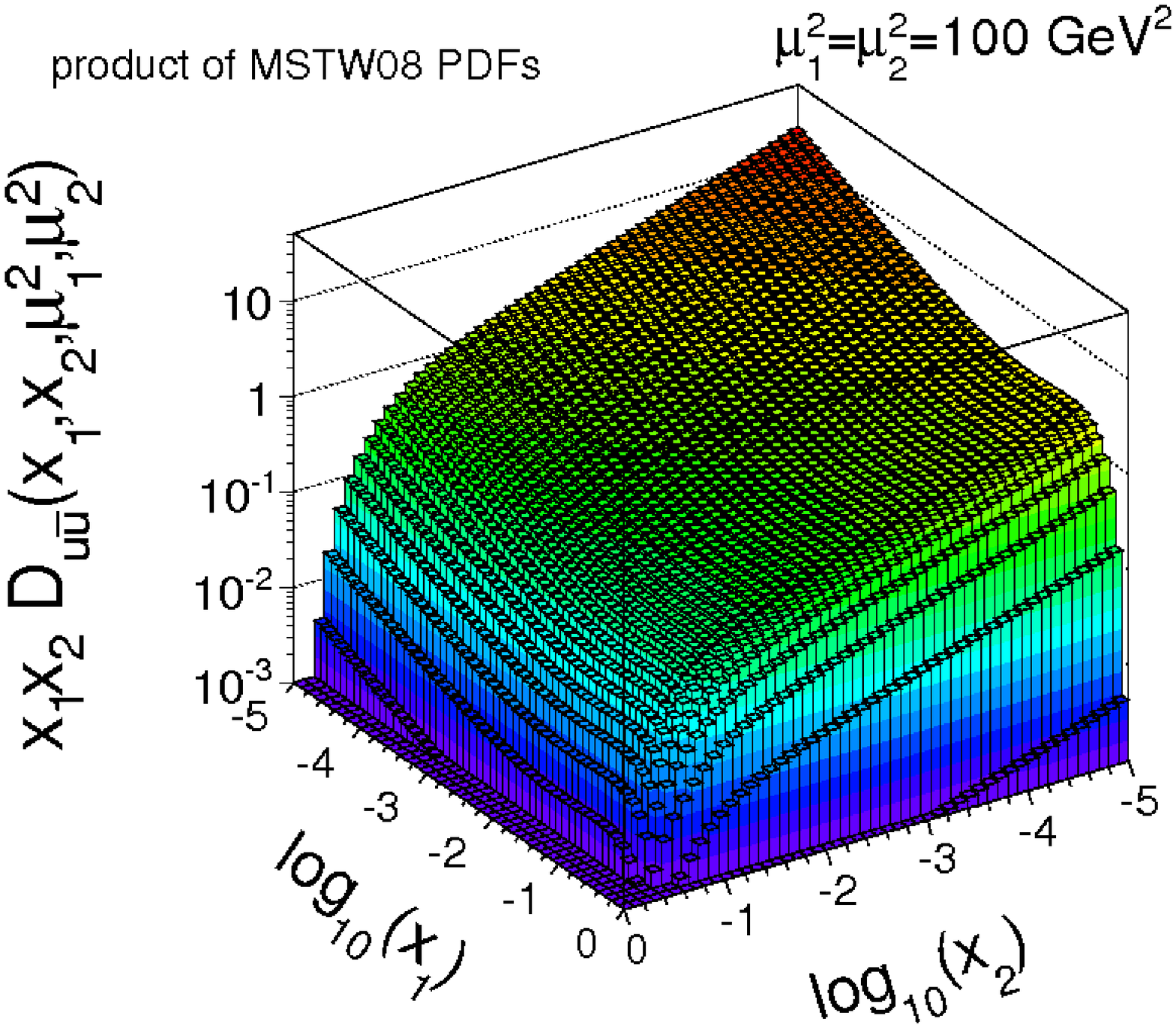}
\includegraphics[width=7.3cm]{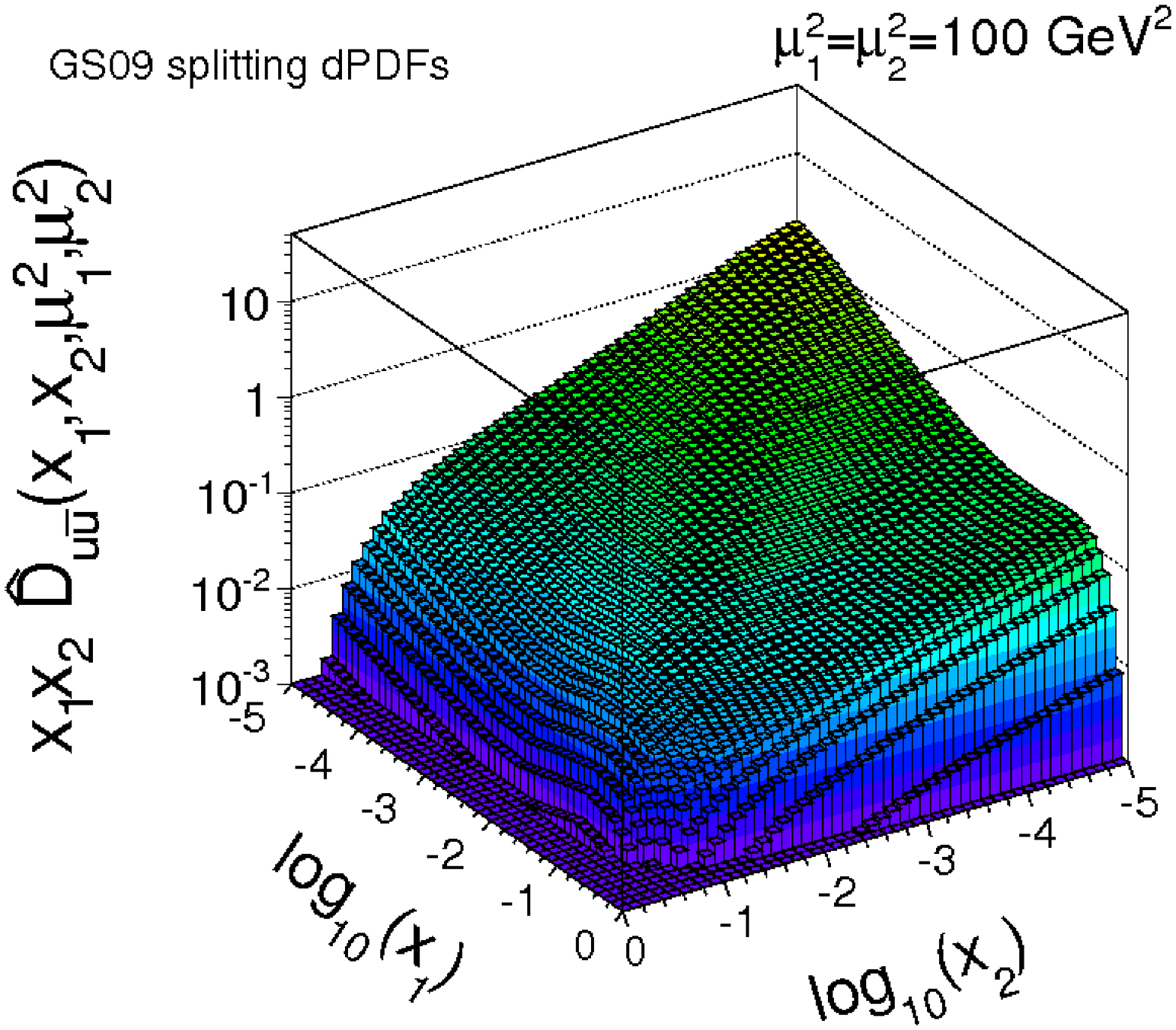} \\
   \caption{
\small Double parton distribution functions:
standard (left column) and for perturbative splitting (right column)
for three different parton combinations for $\mu^2$ = 100 GeV$^2$.
}
 \label{fig:D-functions}
\end{figure}

Before we present results for physical processes discussed in the
present paper we wish to compare our independent ladder pair and 
ladder splitting dPDFs, $D$ and $\hat D$. In Fig.~\ref{fig:D-functions} 
we show plots of the dPDFs for selected parton combinations, and with
factorization scale $\mu^2$ = 100 GeV$^2$ (this is relevant for instance 
for $\chi_b$ meson production). The dPDFs shown are representative
for all (49) combinations included in our full analysis. 

\begin{figure}[!h]
\includegraphics[width=8cm]{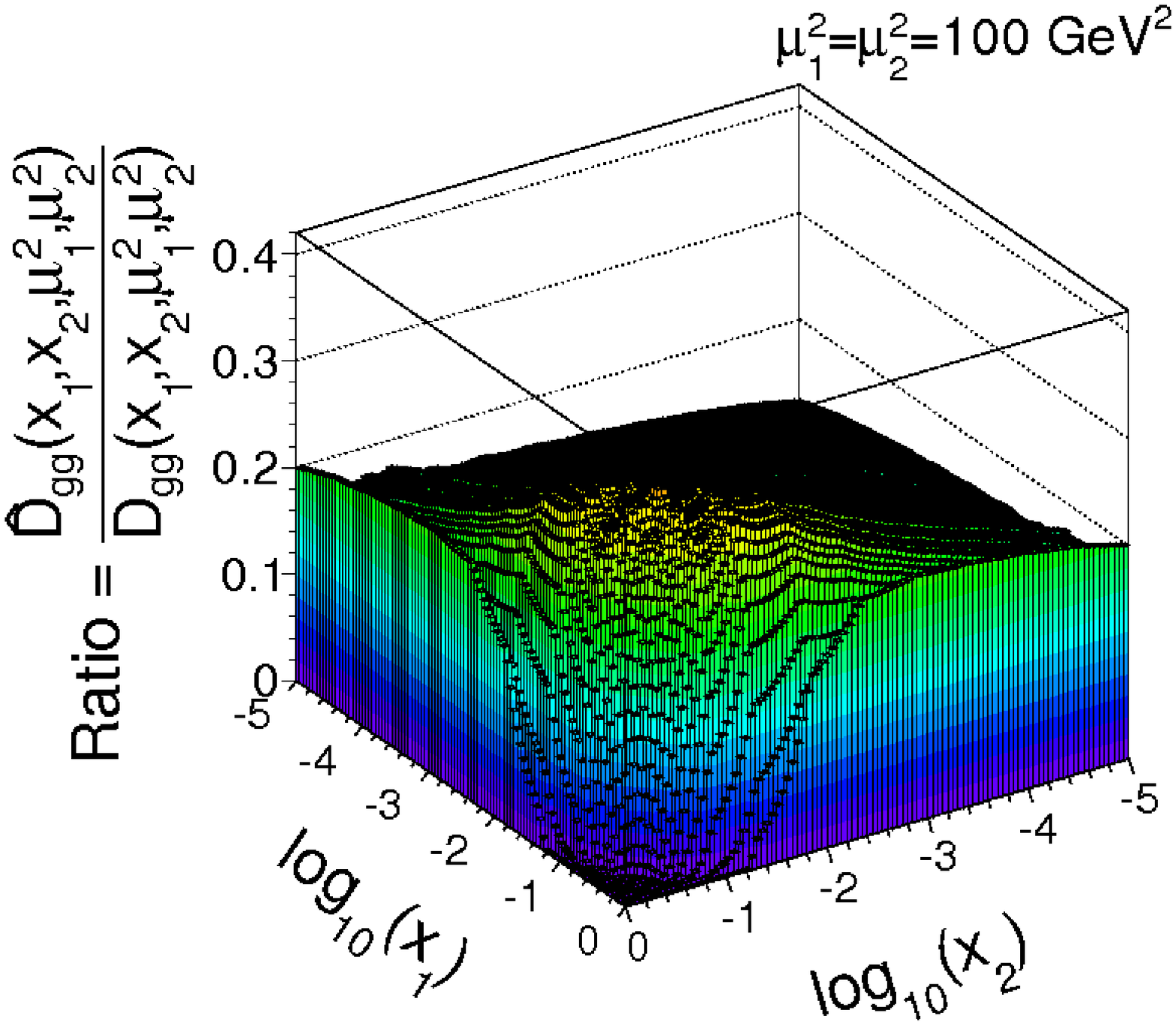}
\includegraphics[width=8cm]{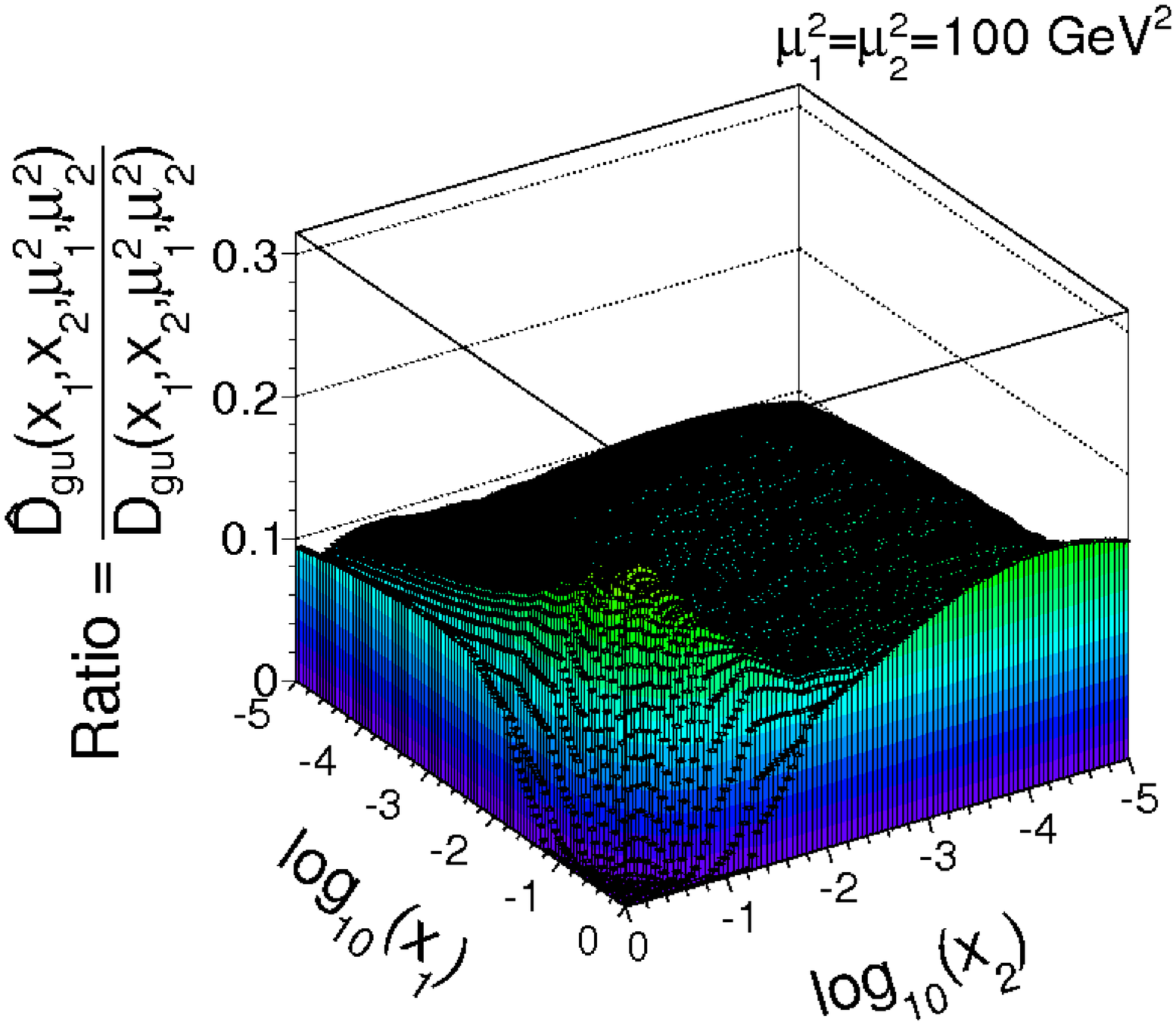} \\
\includegraphics[width=8cm]{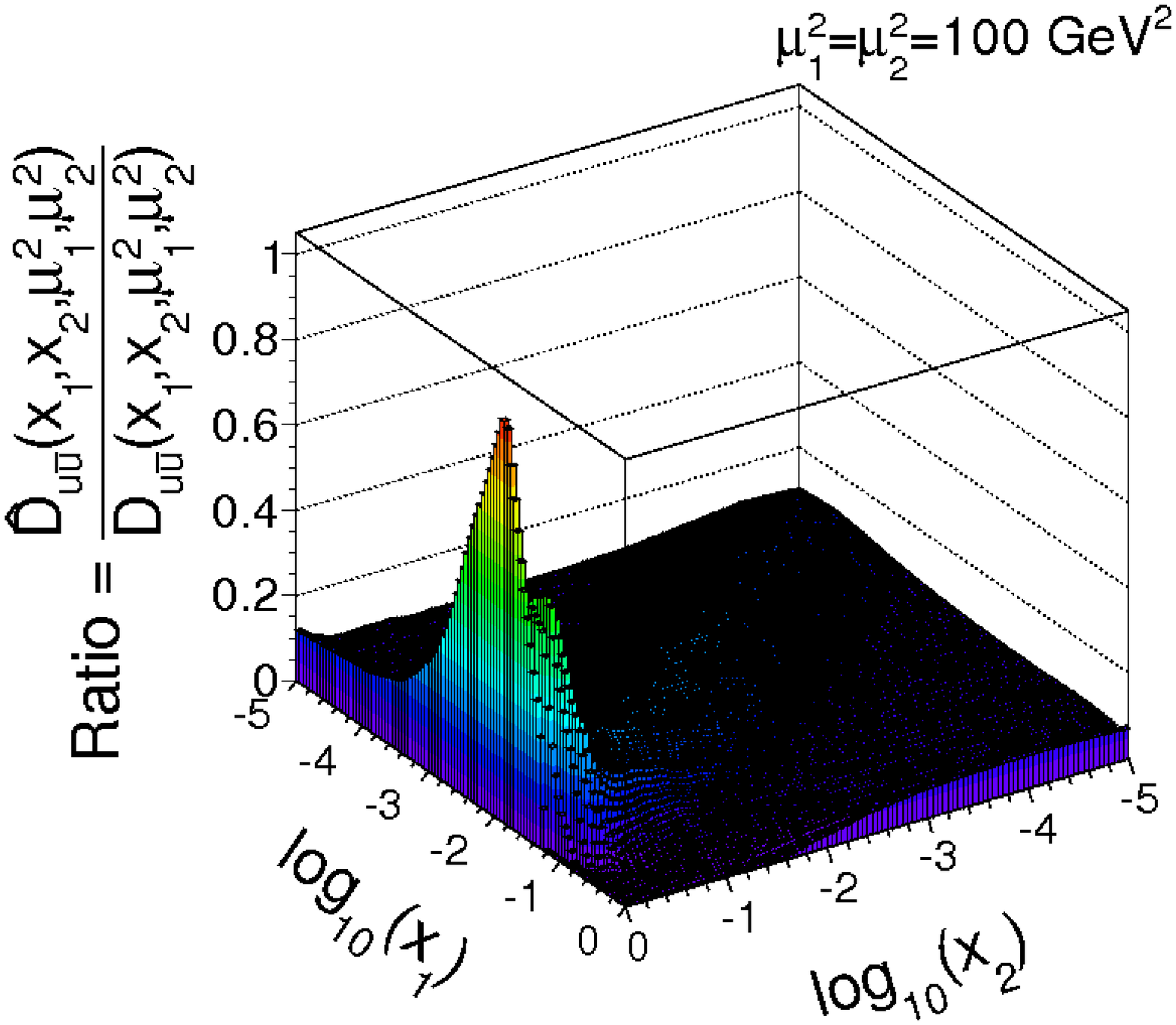}
\includegraphics[width=8cm]{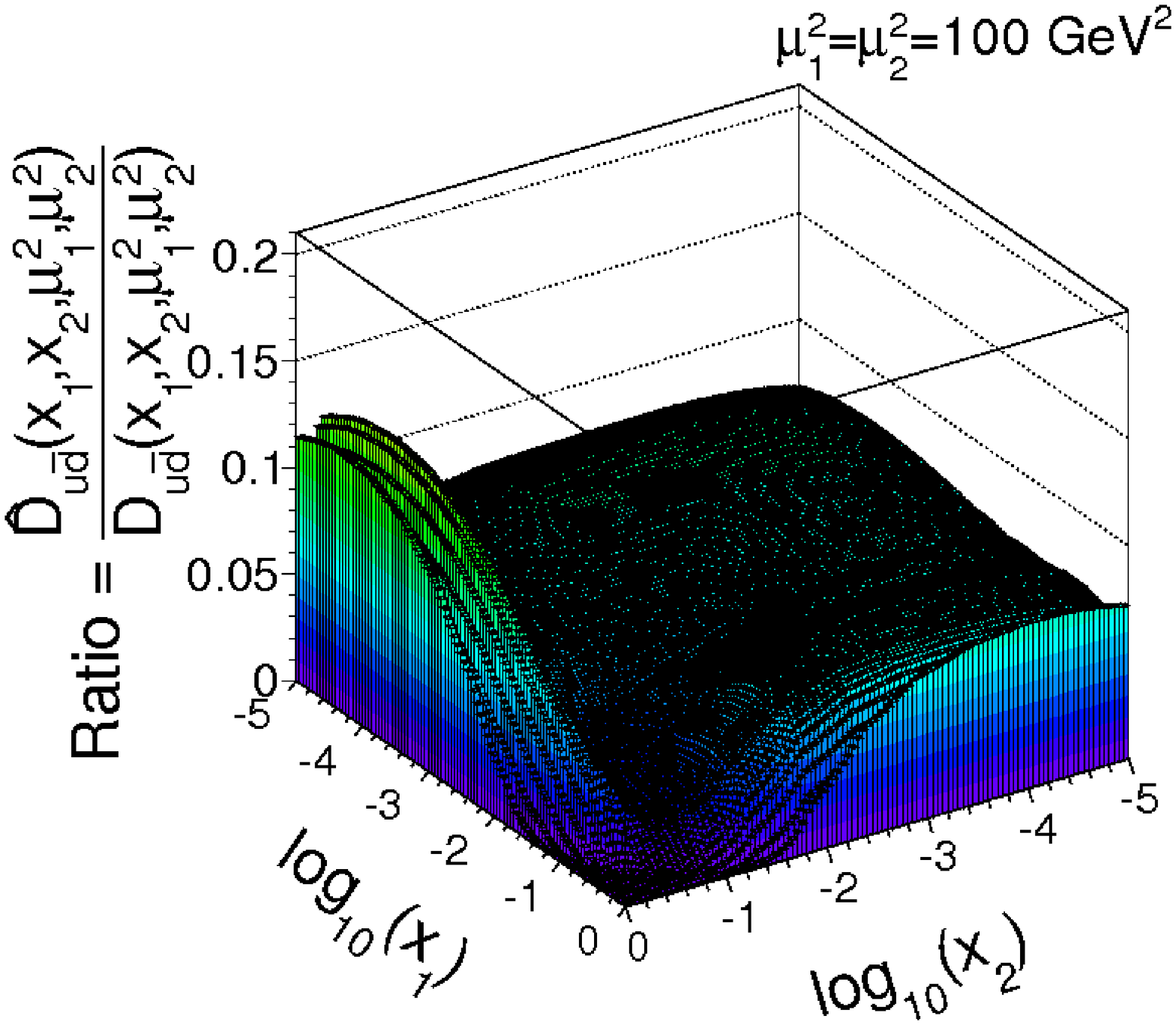}
   \caption{
\small Ratios of perturbative splitting to conventional
double parton distributions for $gg$ (top left), $gu$ (top right),
$u \bar u$ (bottom left) and $u\bar{d}$ (bottom right).
}
 \label{fig:DtoD_ratios}
\end{figure}

One sees that the shapes of the dPDFs differ for the different parton
combinations. Also, the overall size of the ladder splitting dPDFs are
rather smaller than the independent ladder pair dPDFs -- $\hat D/D$ is
typically of order $10\%$ at small $x_1,x_2$. However, one notices that
the shapes of the ladder splitting and independent ladder pair dPDFs are
rather similar for fixed parton flavours $ij$, at least by eye. To get a
better quantitative handle on this, we have plotted the ratios for each
representative parton combination in Fig.~\ref{fig:DtoD_ratios}. Indeed
we see that the ratio takes a roughly constant value of $10\%$ for small
$x_1,x_2$. This is in accord with the plots of $\hat D/D$ (or $1 -\hat D/D$)
along the line $x_1 = x_2$ given in 
\mycites{Korotkikh:2004bz, Cattaruzza:2005nu, GS2010} 
(although note that these plots were produced in the context of the 
old framework of \mycites{K79, SSZ82,S2003}). 

We believe that this similarity
in shapes for small $x_1,x_2$ is related to the observation made in 
\mycites{Gaunt:2012dd, RS2012} that for small $x_1,x_2$ the $1 \to 2$ 
splitting in $\hat D$ typically occurs extremely `early' in $\mu$ 
(just above $Q_0$ -- e.g. less than 3 GeV for $Q=10$ GeV  even 
for rather large $x$ values of order $10^{-1}$ \cite{Gaunt:2012dd}). 
Then, over most of the evolution range, the dominant evolution for the 
$\hat{D}$ is the same as that for the $D$ (i.e. two parton branching 
evolution), and the similar evolution for $D$ and $\hat{D}$ is what 
causes their shapes to converge. In order to test this idea we used 
the numerical code of \mycite{GS2010} to calculate $D$ at $Q = 10$ GeV, 
taking various different forms for the input $D$ at $Q_0 = 1$ GeV 
(a constant, $(1-x_1-x_2)$, $x_1^{-a}x_2^{-a}(1-x_1-x_2)$ with
$a=0.5$ or $1$, etc.). For simplicity we set all the $D^{ij}$s to be 
the same -- in practice the input $D^{gg}$ is the important one 
determining the size of the $D$s at low $x_1,x_2$. 
We found very similar shapes in $D$ for $Q = 10$ GeV and
$x_1,x_2 \lesssim 10^{-2}$ regardless of the input distribution, which 
supports the idea that it is the evolution that causes the shapes 
to be similar. This qualitative behaviour is also found analytically 
in the double leading logarithmic approximation to the parton 
distributions \cite{Ellis:1991qj}, which is supposed to be valid in 
the limit $Q^2 \to \infty , x \to 0$. In this approximation
one finds that the low $x$ behaviour is built up from the perturbative 
evolution, provided that the starting distribution is not too steep.

Another feature of note in the 
ratio plots is the large enhancement of the $u\bar{u}$ ratio when the $x$ 
fraction of the $\bar{u}$ is close to $1$, and the $x$ fraction of the $u$
is not too small -- between $10^{-3}$ and $10^{-1}$. The ratio is large here because
the independent splitting dPDF is suppressed by the small size of the $\bar{u}$
single PDF factor, whilst the perturbative splitting dPDF receives comparatively 
large contributions from direct $g \to u\bar{u}$ splittings (the $g$ that splits
then has to have a rather large $x$, but the MSTW2008LO gluon density is 
quite large at $\mu^2$ = 100 GeV$^2$ even at large $x$). As the $x$ fraction
of the $u$ is decreased, the contribution from direct $g \to u\bar{u}$ splittings
to $\hat D$ remains similar (since in this region it only depends on the much 
larger $x$ of the $\bar{u}$), whilst the independent pair dPDF increases due to the
$u$ PDF factor, and the ratio decreases. This explanation can be tested by plotting
the ratio for the parton combination $u\bar{d}$ -- then we expect no enhancement in 
the ratio of the kind that we found for the $u\bar{u}$. This is because a gluon cannot
directly split into a $u\bar{d}$ pair. We include the $u\bar{d}$ ratio as the final
plot in Fig.~\ref{fig:DtoD_ratios}, and indeed find that no enhancement of the ratio
for this plot is found.

A further interesting point to make about the $u\bar{d}$ plot is that the ratio is 
roughly the same as the $gg$, $ug$ or $u\bar{u}$ at small $x_1,x_2$ even though this 
distribution receives no direct feed from the inhomogeneous term in \eqref{dbDGLAP}.
This is due to the aforementioned point that for small final $x_1,x_2$ the $1 \to 2$ splitting
occurs very early, leaving plenty of evolution space for further emissions that allow
(for example) a $g$ to eventually give rise to a $u\bar{d}$ pair (plus various other
emitted partons). This means that we cannot suppress the 2v1 contribution to DPS by
picking processes such as same sign $WW$ that are initiated by such parton pairs
(unless one finds a way to probe very large $x$s in this process).

The similar shape of the ladder splitting dPDFs for small $x_1,x_2$ as compared 
to the independent ladder dPDFs indicates that the differential cross section 
contributions associated with the 2v1 and 2v2 mechanisms will be rather similar.
This we will see in the next two subsections.

%----------------------------------------------------------------
%----------------------------------------------------------------

%----------------------------------------------------------------
%----------------------------------------------------------------

%--------------------------------------------
\subsection{Quarkonium production}
%---------------------------------------------

In the calculations, results of which will be discussed below, we assume 
$\mu_1^2, \mu_2^2 = M_{\chi}^2$, where $M_{\chi}$ is a generic name for
the S-wave, P-wave quarkonium or Higgs boson mass.

In Table~\ref{table:ratio} we present the ratio of 
$\sigma^{2v1}/\sigma^{2v2}$ for the production of two identical-mass
objects (two identical quarkonia, two Higgs bosons).
Following our earlier discussion from section \ref{sec:DPSsetup}
we take the ratio $\sigma_{eff,2v2}/\sigma_{eff,2v1} = 2$.
The ratio only slightly depends on the mass of the object and
centre-of-mass energy but the tendency is rather clear.
The masses chosen correspond roughly to production of 
$\eta_c$, $\chi_c$ (M = 3 GeV), $\eta_b$, $\chi_b$ (M = 10 GeV)
quarkonia and Higgs boson (M = 126 GeV).
The double Higgs case is purely academic as the corresponding DPS
cross section is rather small (a $\sim$ 10$^{-4}$ fraction of fb,
much smaller than the single parton scattering cross section 
\cite{Plehn:1996wb, Dawson:1998py, Binoth:2006ym, Baglio:2012np, Frederix:2014hta}) 
but the effect of the perturbative splitting can be here well 
illustrated.

%--------------------------------------------------------------------------
\begin{table}
\caption{The ratio of $\sigma^{2v1}/\sigma^{2v2}$ for double
quarkonium production (full phase space) for different masses 
of the produced object (rows)
and different centre-of-mass energies (columns) in TeV.
}
\label{table:ratio}
\begin{tabular}{|c|c|c|c|c|c|}
\hline
%\diaghead{ xxxxxxxxxxxxxx}%
{M (GeV)} / {$\sqrt{s}$ (TeV)}& 0.2   & 0.5   & 1.96  & 8.0   & 13.0 \\
%M (GeV) / $\sqrt{s}$ (TeV) & 0.2   & 0.5   & 1.96  & 8.0   & 13.0  \\
\hline
3.                   & 0.840  & 0.775  & 0.667  & 0.507  & 0.437  \\ 
10.                  & 1.116  & 1.022  & 0.891  & 0.780  & 0.743  \\
126.                 & --    &  ---  & 1.347  & 1.134  & 1.070  \\  
\hline
\end{tabular}
\end{table}
%-----------------------------------------------------------------------------

\begin{figure}[!h]
\includegraphics[width=7.5cm]{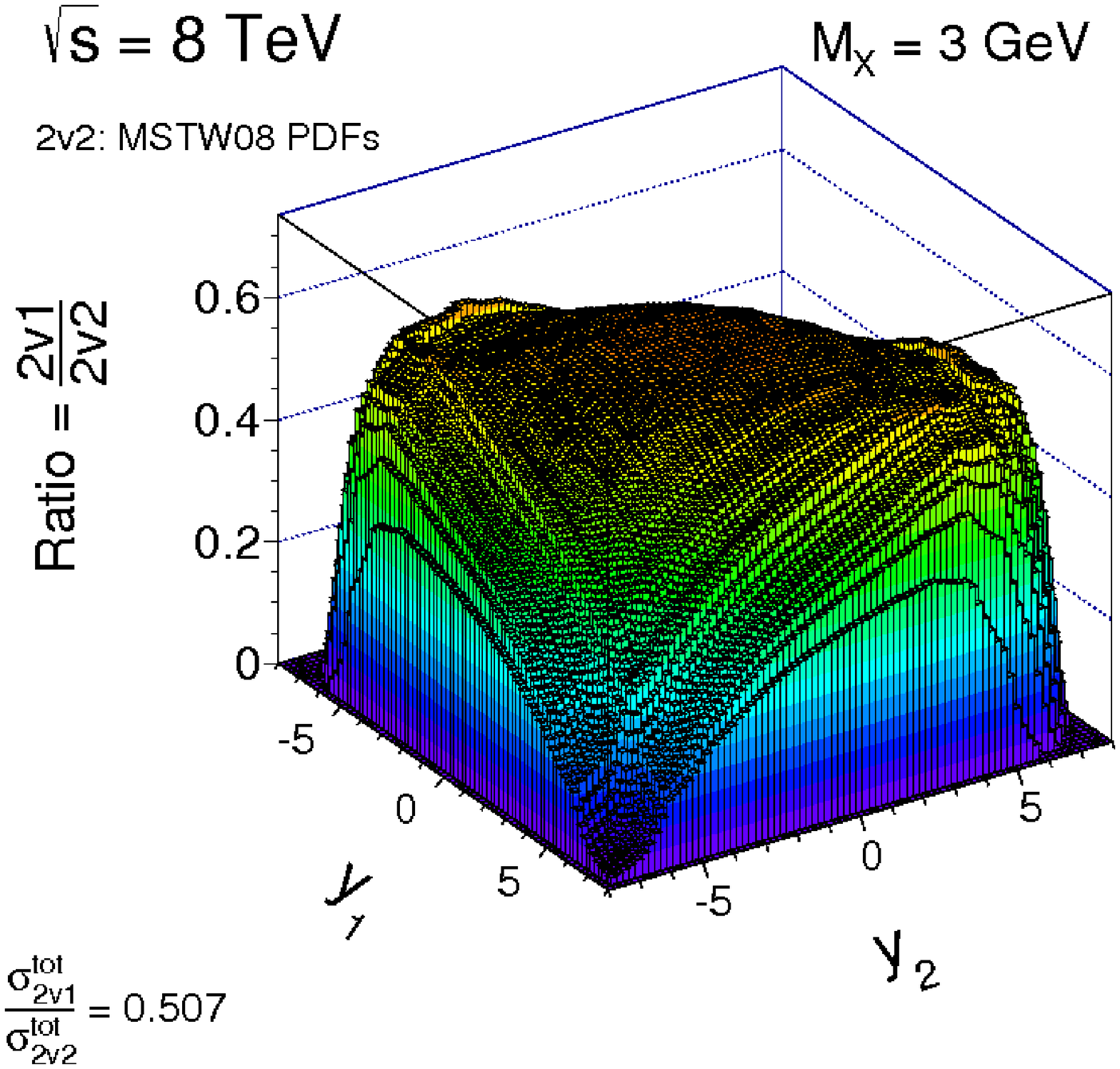}
\includegraphics[width=7.5cm]{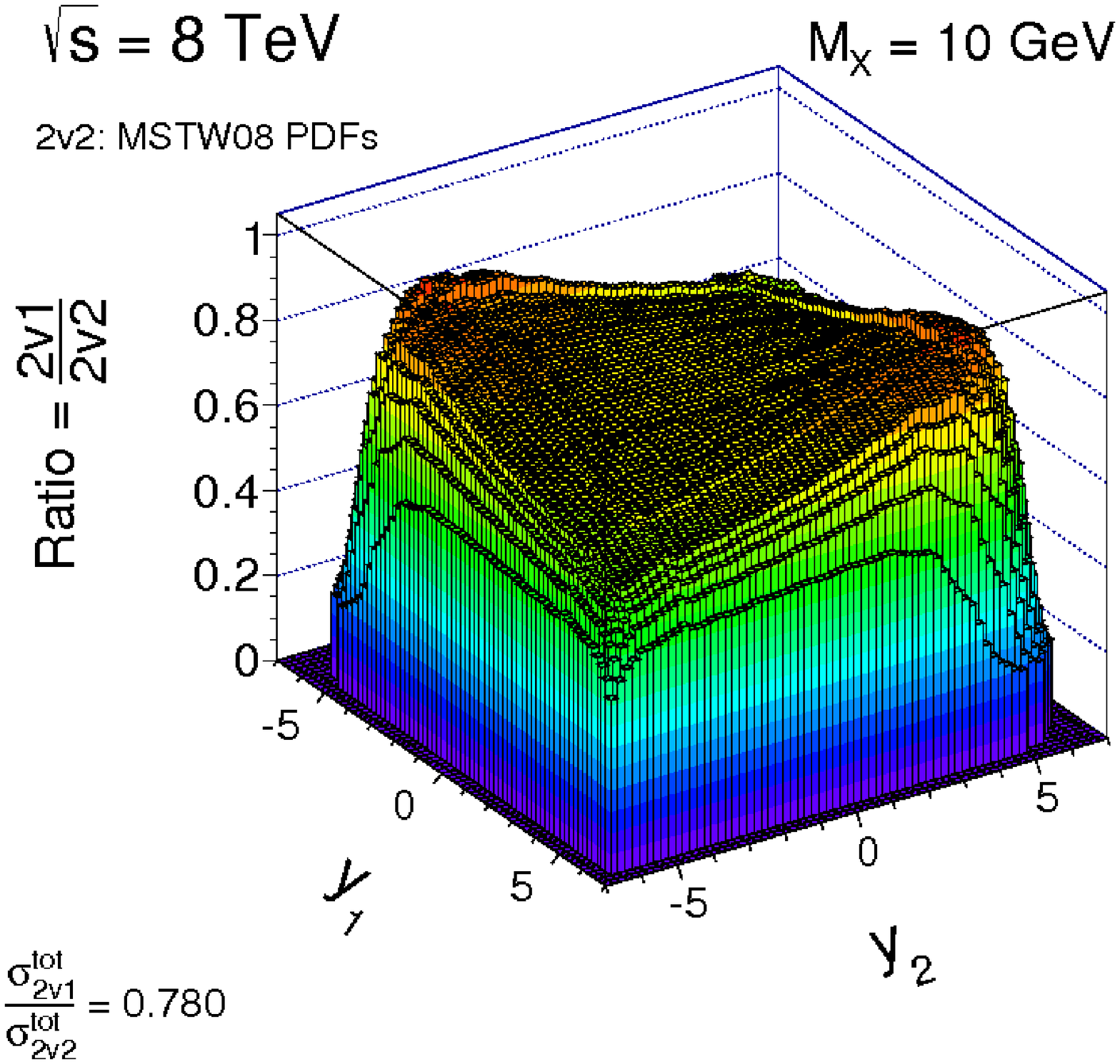}
\includegraphics[width=7.5cm]{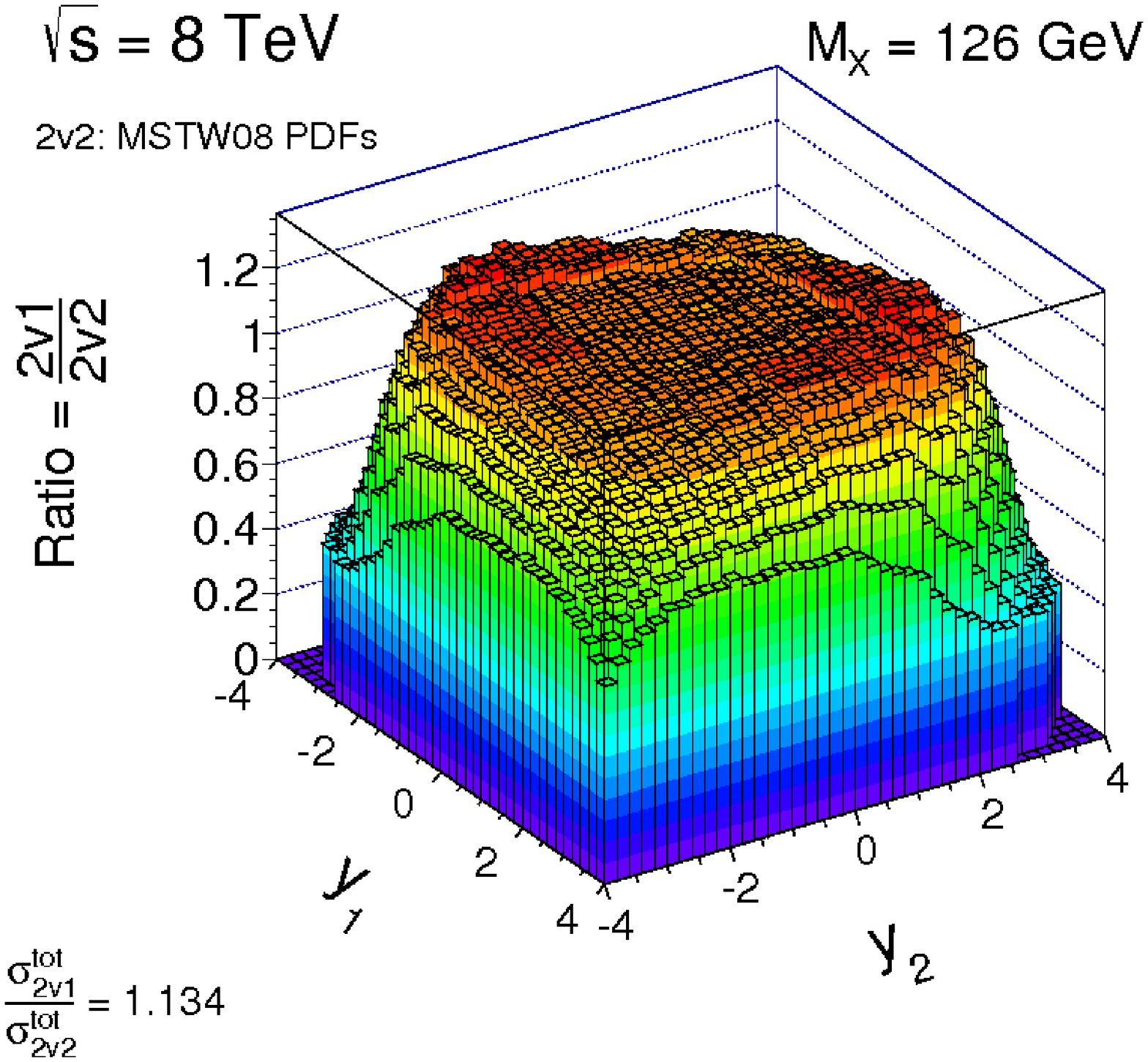}
   \caption{
\small $R(y_1,y_2)$ for $\sqrt{s}$ = 8 TeV for different masses:
M = 3 GeV (top-left), M = 10 GeV (top-right) and M = 126 GeV (bottom-middle).
}
 \label{fig:y1y2}
\end{figure}

In Fig.~\ref{fig:y1y2} we show the ratio defined as
\begin{equation}
R(y_1,y_2) = 
\frac{\frac{d \sigma^{2v1}}{dy_1 dy_2}(y_1,y_2)}
{\frac{d \sigma^{2v2}}{dy_1 dy_2}(y_1,y_2)} \; .
\label{ratio}
\end{equation}

From these plots we can see that $R(y_1,y_2)$ does not depend strongly
on the rapidities $y_1$ and $y_2$, as one would expect given that the 
ratio of the ladder splitting and independent ladder pair dPDFs does not
depend strongly on $x_1,x_2$ (Fig.~\ref{fig:DtoD_ratios}).

%----------------------------------------------------------------
%----------------------------------------------------------------

In the calculation of the cross sections in this and in the next 
subsection we have to fix the two nonperturbative parameters:
$\sigma_{eff,2v2}$ and $\sigma_{eff,2v1}$. Their values are not well
known. Once again we take the ratio $\sigma_{eff,2v2}/\sigma_{eff,2v1} =
2$. We choose $\sigma_{eff,2v2}$ = 30 mb which corresponds to assuming 
that partons in a `nonperturbatively generated' pair are essentially 
uncorrelated in transverse space \cite{Frankfurt:2003td, Calucci:1999yz}
(but note that varying $\sigma_{eff,2v2}$ with the ratio 
$\sigma_{eff,2v2}/\sigma_{eff,2v1}$ fixed affects only the
normalisations of the cross sections presented below).

In Fig.~\ref{fig:sig_eff_quarkonia} we show how the empirical 
$\sigma_{eff}$ value depends on centre-of-mass energy assuming that 
the value of $\sigma_{eff,2v2}$ is independent of energy. 
We see a clear dependence of $\sigma_{eff}$ on energy in the plot, and also
on the mass of the quarkonium. Assuming that there is no other mechanism
for an energy dependence of $\sigma_{eff}$, $\sigma_{eff}$ is 
therefore expected to increase with centre-of-mass energy. 
Note also that the empirical $\sigma_{eff}$ value obtained is 
in the ball park of the values extracted in experimental
measurements of DPS ($\sim 15$ mb), even though $\sigma_{eff,2v2}$ is 
rather larger, assumed here to be 30 mb.

%----------------------------------------------------------------
\begin{figure}[!h]
\includegraphics[width=8cm]{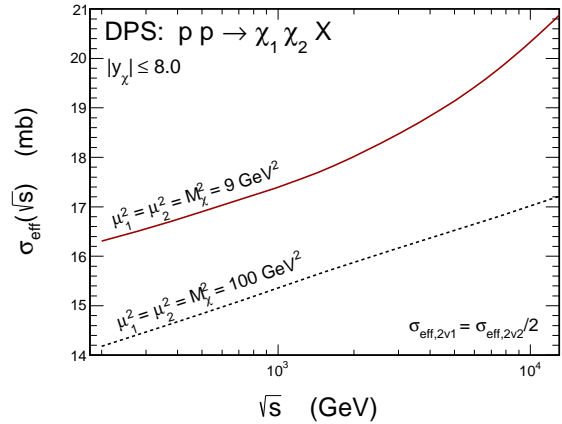}
   \caption{
\small Energy and quarkonium mass dependence of $\sigma_{eff}$
as a consequence of existence of two components. In this calculation
we have taken $\sigma_{eff,2v2}$ = 30 mb and $\sigma_{eff,2v1}$ = 15 mb.
}
 \label{fig:sig_eff_quarkonia}
\end{figure}
%----------------------------------------------------------------

%-------------------------------------------------
\subsection{$\bm{c \bar c c \bar c}$ production}
%-------------------------------------------------

Now we proceed to double charm production. 
Here we either assume $\mu_1^2 = m_{1t}^2$ and $\mu_2^2 = m_{2t}^2$,
or $\mu_1^2 = M_{c\bar c,1}^{2}$ and $\mu_2^2 = M_{c\bar c,2}^{2}$.
The quantity $m_{it}$ is the transverse mass of either parton emerging
from subprocess $i$, whilst $M_{c \bar c,i}$ is the invariant mass of 
the pair emerging from subprocess $i$.
In Table \ref{table:ratio_ccbarccbar} we show the ratio of 2v1-to-2v2
cross sections for different centre of mass energies.
The numbers here are similar to those for the double quarkonium production.

%--------------------------------------------------------------------------
\begin{table}
\caption{The ratio of $\sigma^{2v1}/\sigma^{2v2}$ for $c \bar c c \bar c$
production (full phase space) for different centre-of-mass energies in TeV.
}
\label{table:ratio_ccbarccbar}
\begin{tabular}{|c|c|c|c|c|c|}
\hline
$\mu^{2}$ (GeV$^2$) / $\sqrt{s}$ (TeV) & 0.2    & 0.5    & 1.96   & 7.0    & 13.0  \\
\hline 
$m_{t}^{2}$            & 0.628  & 0.610  & 0.503  & 0.326  & 0.254  \\ 
$M_{c\bar c}^{2}$      & 0.914  & 0.855  & 0.760  & 0.667  & 0.606 \\
\hline

\end{tabular}
\end{table}
%-----------------------------------------------------------------------------

\begin{figure}[!h]
\includegraphics[width=6.5cm]{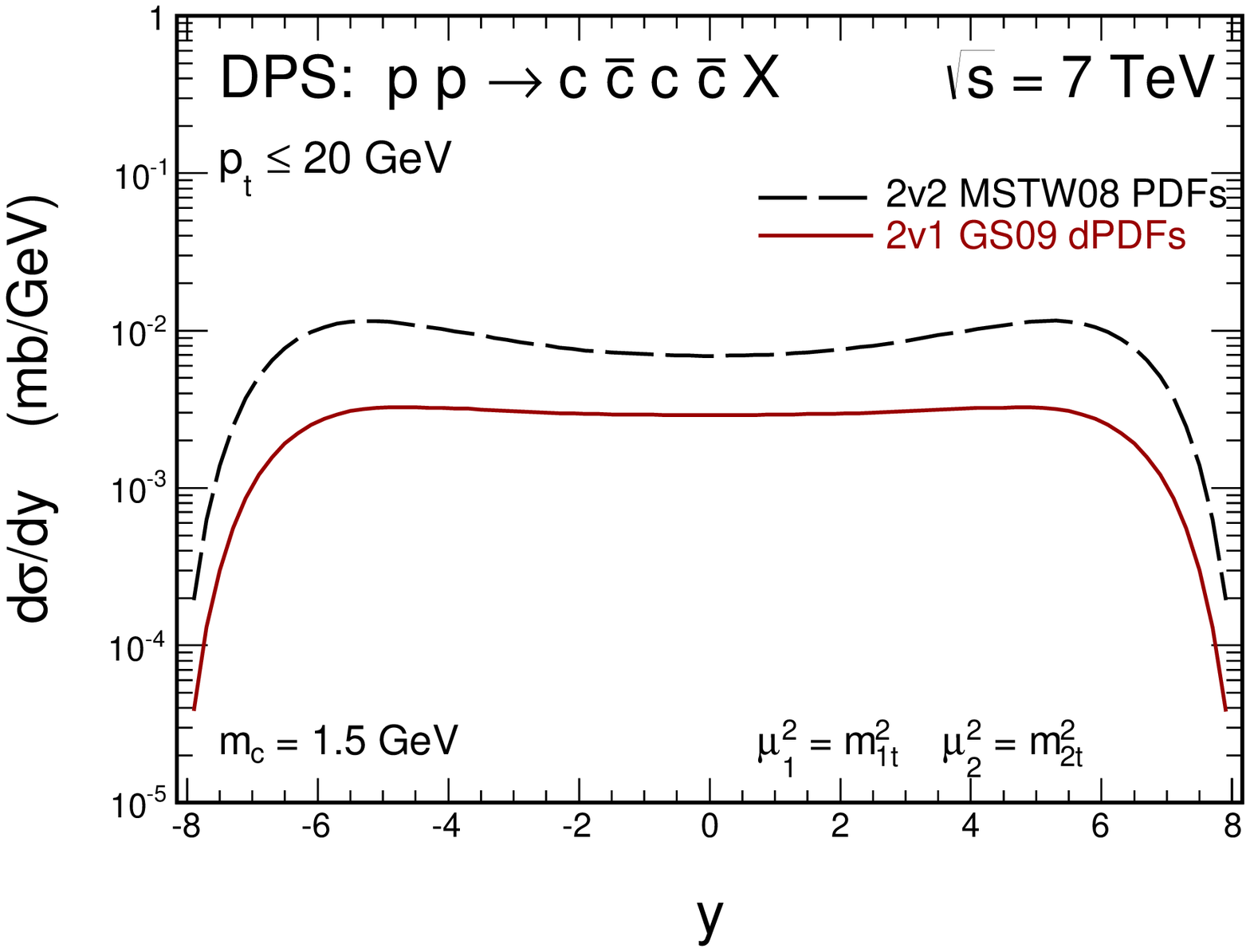}
\includegraphics[width=6.5cm]{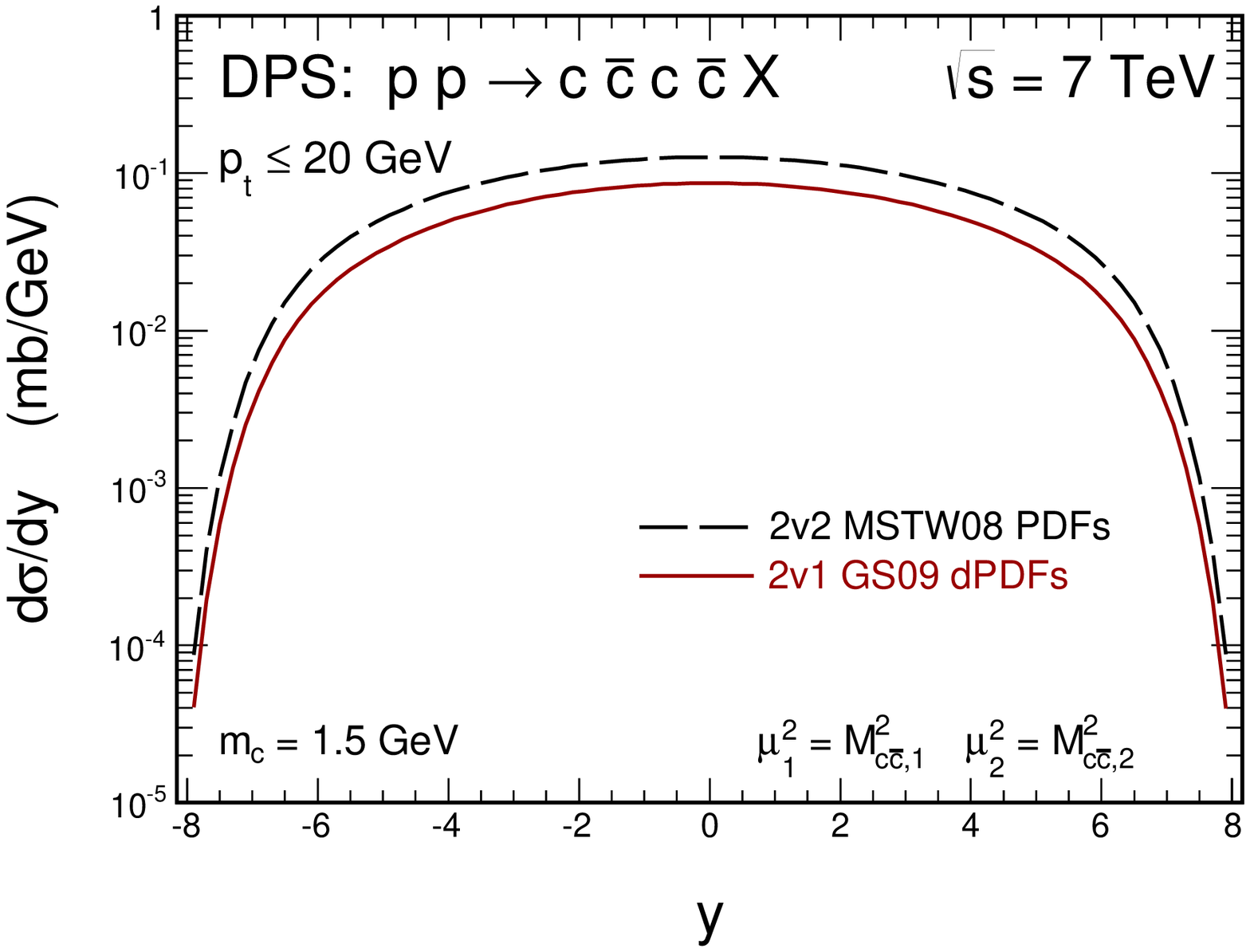}
   \caption{
\small Rapidity distribution of charm quark/antiquark 
for $\sqrt{s}$ = 7 TeV for two different choices of scales:
$\mu_1^2 = m_{1t}^2$, $\mu_2^2 = m_{2t}^2$ (left) and
$\mu_1^2 = M_{c \bar c,1}^2$, $\mu_2^2 = M_{c \bar c,2}^2$ (right).  
}
 \label{fig:dsig_dy}
\end{figure}

Let us show now some examples of differential distributions.
In Fig.~\ref{fig:dsig_dy} we show the rapidity distribution of the charm
quark/antiquark for different choices of the scale at $\sqrt{s}$ = 7 TeV. 
The conventional and splitting terms are shown separately. 
The splitting contribution (lowest curve, red online) is smaller, 
but has almost the same shape as the conventional DPS contribution. 
%The conventional contribution is shown for different approximations 
%as explained in the figure caption. The different approximations 
%give rather similar result.
We wish to note the huge difference arising from the different choices 
of factorization scale. The second choice $\mu^2 = M_{c \bar c}^2$ leads 
to cross sections more adequate for the description of the LHCb data
for double same-flavor $D$ meson production \cite{LHCb}.

%----------------------------------------------------------------
%----------------------------------------------------------------

\begin{figure}[!h]
\includegraphics[width=6.5cm]{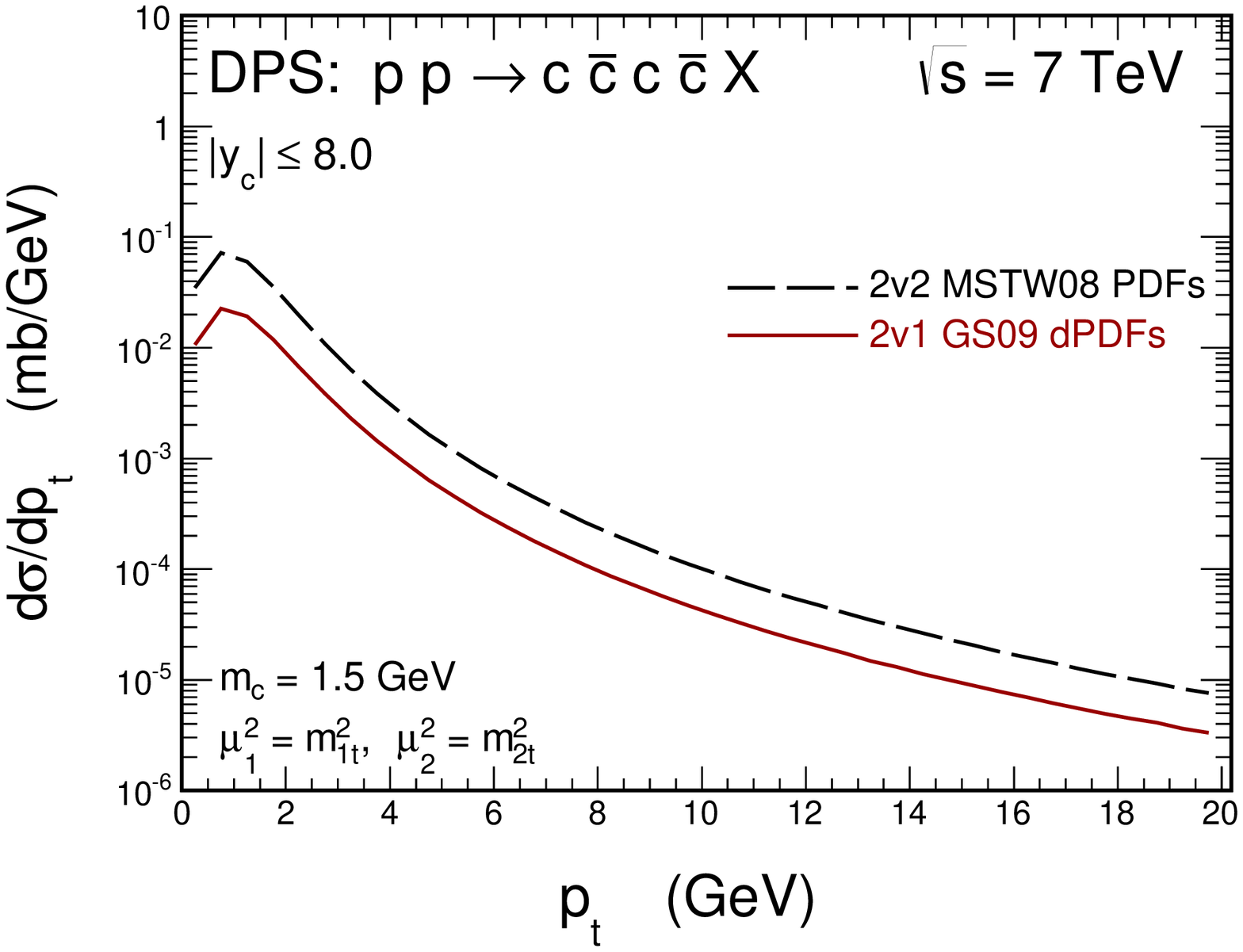}
\includegraphics[width=6.5cm]{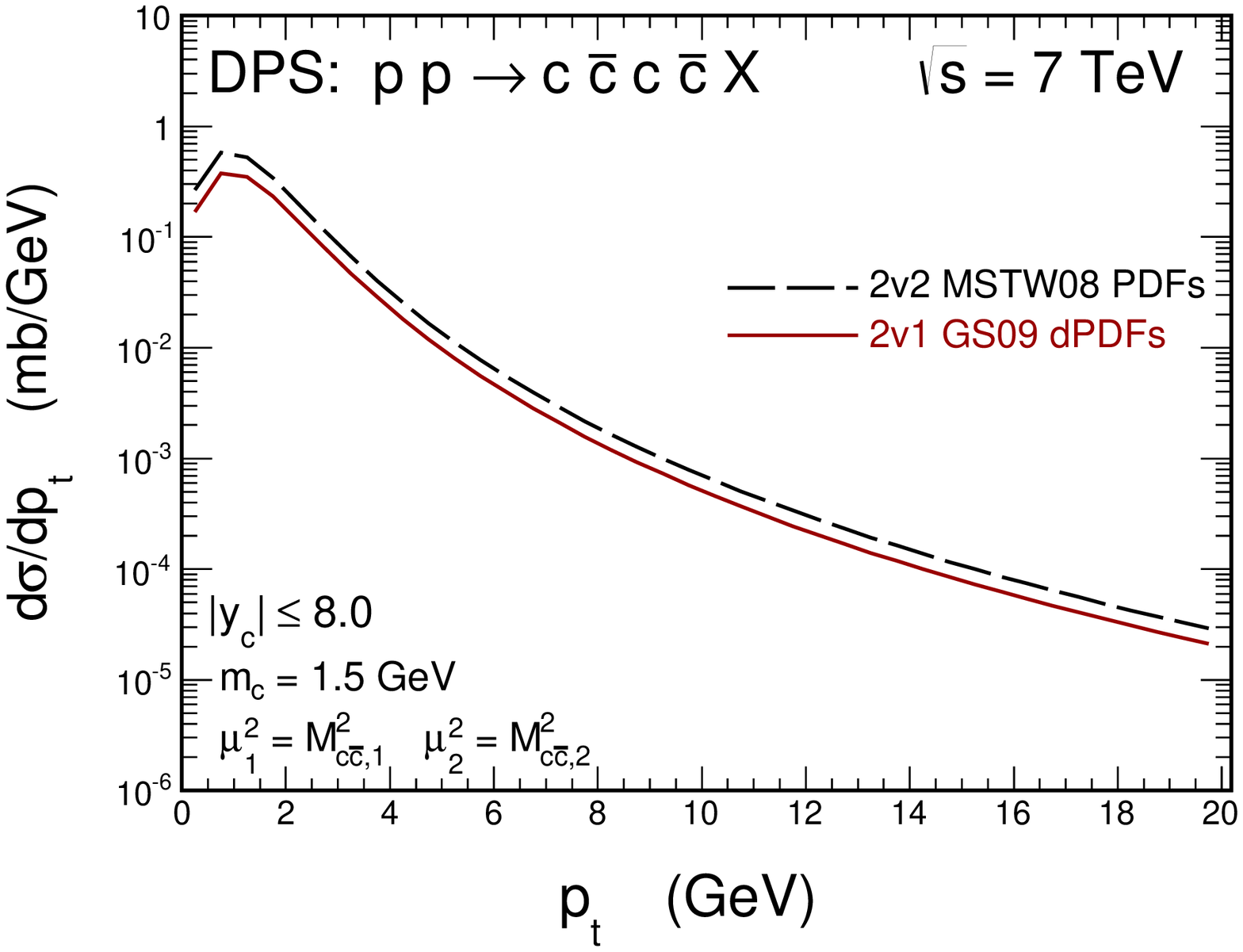}
   \caption{
\small Transverse momentum distribution of charm quark/antiquark 
for $\sqrt{s}$ = 7 TeV for two different choices of scales. 
}
 \label{fig:dsig_dpt}
\end{figure}

In Fig.~\ref{fig:dsig_dpt} we show corresponding distributions
in transverse momentum of charm quark/antiquark. Again the shapes
of conventional and splitting contributions are almost the same.

%----------------------------------------------------------------
%----------------------------------------------------------------

\begin{figure}[!h]
\includegraphics[width=6.5cm]{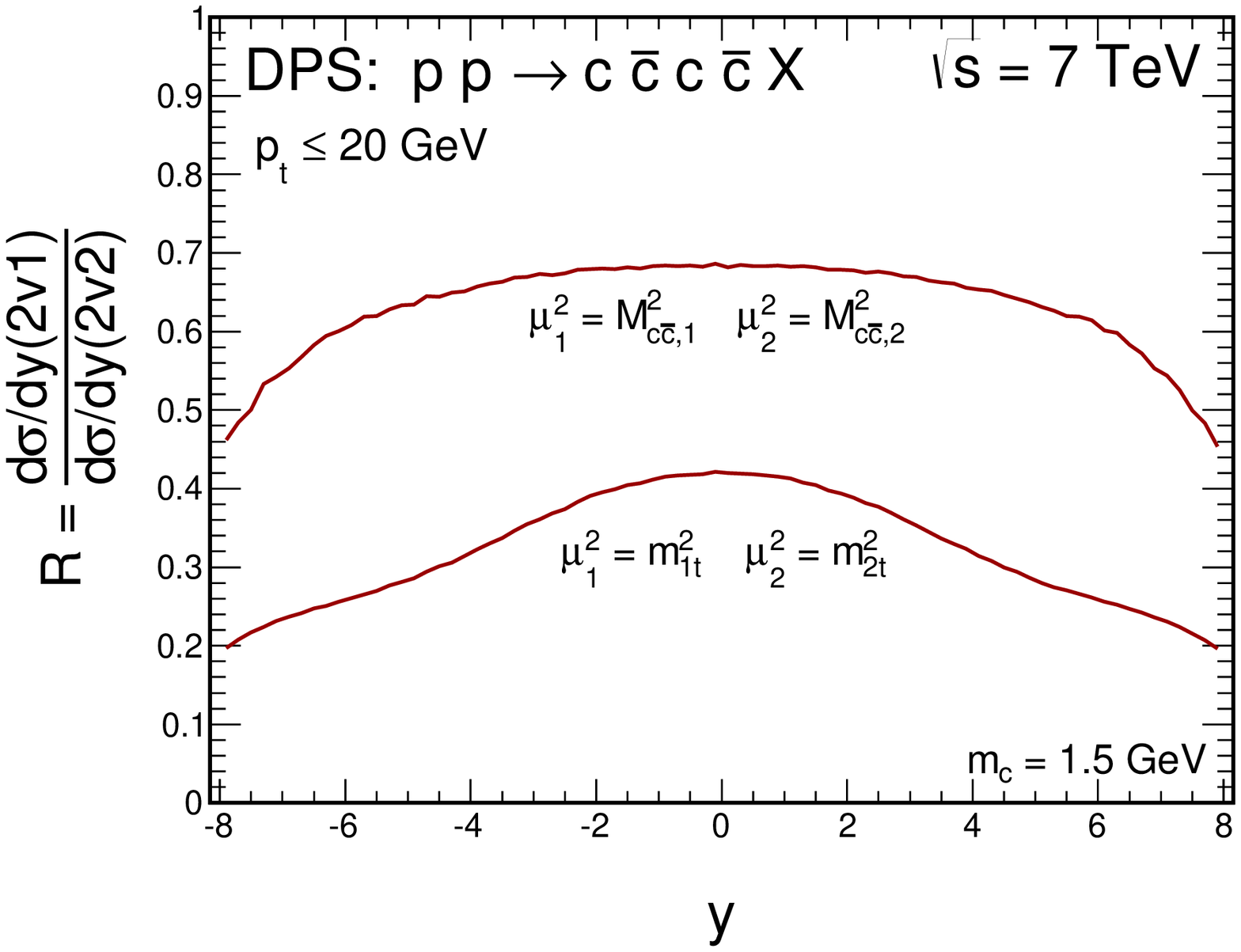}
\includegraphics[width=6.5cm]{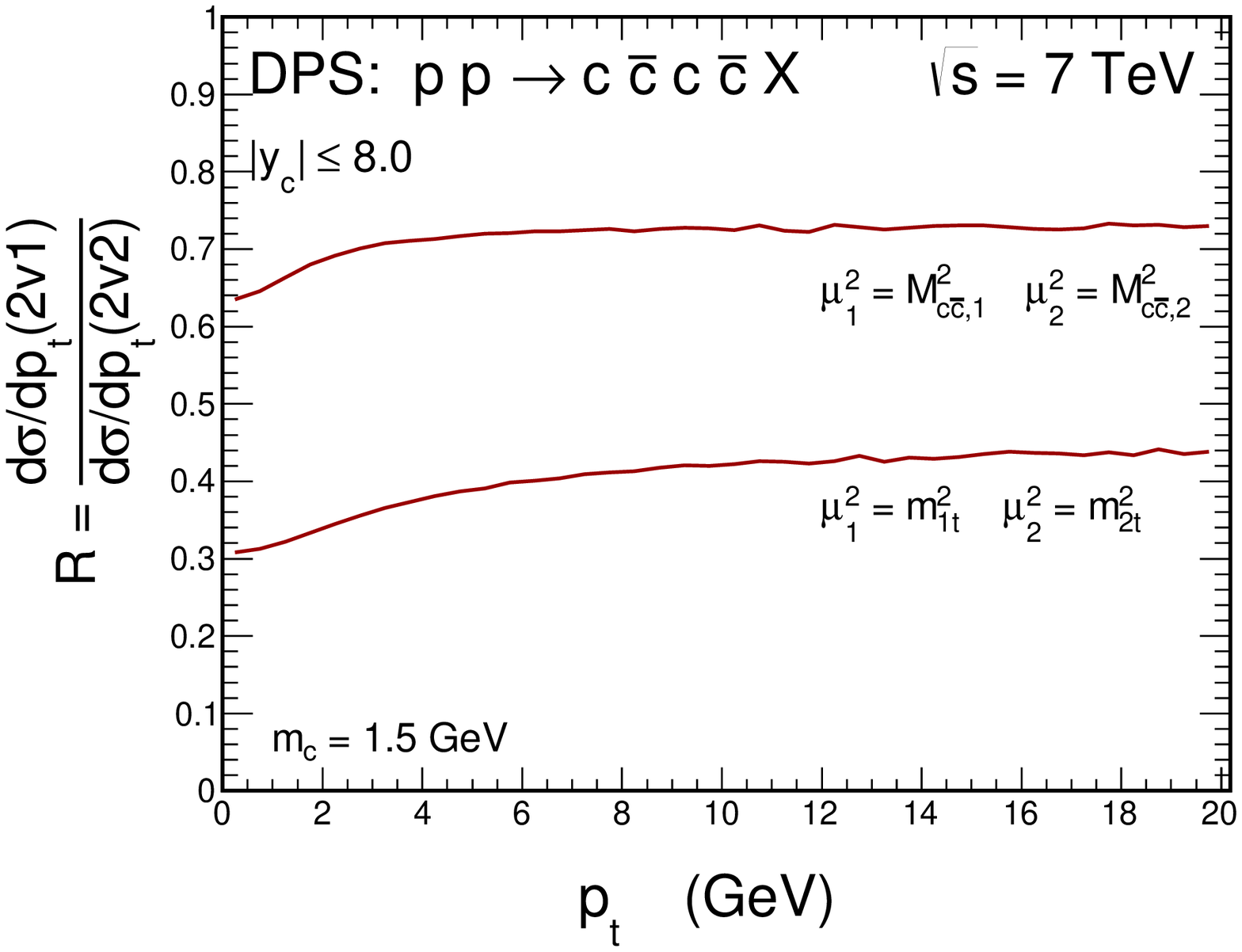}
   \caption{
\small The ratios of 2v1-to-2v2 contributions as a function of
rapidity (left) and transverse momentum (right)
for $\sqrt{s}$ = 7 TeV for two different choices of scales. 
}
 \label{fig:ratios}
\end{figure}

The corresponding ratios of the 2v1-to-2v2 contributions 
as a function of rapidity (left) and transverse momentum (right)
are shown in Fig.~\ref{fig:ratios}. Especially the transverse
momentum dependence shows a weak but clear tendency.

%----------------------------------------------------------------
%----------------------------------------------------------------

\begin{figure}[!h]
\includegraphics[width=8cm]{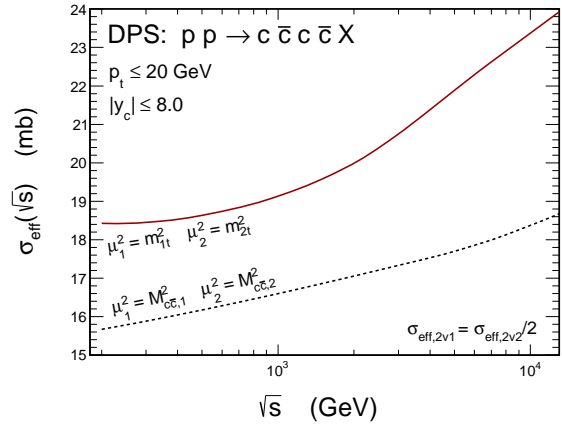}
   \caption{
\small Energy and factorization scale dependence of $\sigma_{eff}$
for $c \bar c c \bar c$ production as a consequence of existence
of the two components. 
In this calculation we have taken
$\sigma_{eff,2v2}$ = 30 mb and $\sigma_{eff,2v1}$ = 15 mb.
}
 \label{fig:sig_eff_charm}
\end{figure}

Finally in Fig.~\ref{fig:sig_eff_charm} we show the empirical $\sigma_{eff}$,
this time for double charm production. Again $\sigma_{eff}$ 
rises with the centre-of-mass energy. A rather large difference
between different choices of scales can be observed.

%----------------------------------------------------------------
%----------------------------------------------------------------

%-------------------------
\section{Conclusion}
%-------------------------

In the present paper we have presented first quantitative estimates 
of the single perturbative splitting 2v1 contribution to
double quarkonium, double Higgs boson and $c \bar c c \bar c$
production. In all cases we find that the splitting contribution
is of the same order of magnitude as the more conventional 2v2
contribution often discussed in the literature. This is consistent with
the observation made already in \mycite{BDFS2013}. 
The perturbative splitting contribution was not considered explicitly
in previous detailed analyses of $c \bar c c \bar c$ and pairs of the 
same-flavor $D$ mesons.

Our calculation shows that the parton-splitting contribution 
is not negligible and has to be included in the full analysis.
However, it is too early in the moment for detailed predictions of the
corresponding contributions as our results strongly depend on the values
of not well known parameters $\sigma_{eff,2v2}$ and $\sigma_{eff,2v1}$.
Both their magnitude and even their ratio are not well known.
We have presented only some examples inspired by a simple geometrical model. 
A better understanding of the two nonperturbative parameters seems 
an important future task.

We have shown that almost all differential distributions (in rapidity,
transverse momentum, even many two-dimensional distributions) 
for the conventional and the parton-splitting contributions 
have essentially the same shape. This makes their 
model-independent separation extremely difficult.
This also shows why the analyses performed so far could describe
different experimental data sets in terms of the conventional 2v2 
contribution alone. The sum of the 2v1 and 2v2 contributions behaves
almost exactly like the 2v2 contribution, albeit with a smaller 
$\sigma_{eff}$ that depends only rather weakly on energy, scale and
momentum fractions.

With the perturbative 2v1 mechanism included, $\sigma_{eff}$ increases as
$\sqrt{s}$ is increased, and decreases as $Q$ is increased. A 
decrease of $\sigma_{eff}$ with $Q$ was also observed in \mycite{Blok:2011bu}
for the same reason. Similar trends were also observed 
in \mycite{Gustaffson}, although the calculation there is performed in a BFKL
framework rather than the DGLAP framework used here. In \mycite{Gustaffson}
the decrease of the effective $\sigma_{eff}$ with $Q$ is somewhat stronger.
It is difficult to pin down the exact reason for this difference due to
the different calculational frameworks used. However, we remark that the 
definitions of the 2pGPDs and total DPS cross section used in \mycite{Gustaffson}
would, in the DGLAP framework, allow some effective 1v1 contribution to
DPS, which here we do not include.
 
At present only the leading-order version of the single perturbative
splitting formalism is available.
However, it is well known that NLO corrections for the gluon initiated 
processes are rather large, also for processes considered here.
In the case of $c \bar c c \bar c$ production they can be taken into 
account e.g. in the $k_t$-factorization \cite{MS2013_ccbar}.
It is not clear in the moment how to combine the higher-order
effects with the perturbative splitting mechanism discussed here.
An interesting question is whether the ratio between the 2v1 and 2v2
contributions changes when higher-order corrections are included. Further studies are clearly needed.

\vspace{1cm}

{\bf Acknowledgments}

One of us (A.S) is indebted to Alexander Snigirev for a discussion
and interesting comments. We thank Markus Diehl for various useful
comments on the manuscript.
This work was partially supported by the Polish NCN grant DEC-2011/01/B/ST2/04535
as well as by the Centre for Innovation and Transfer 
of Natural Sciences and Engineering Knowledge in Rzesz\'ow.

%-----------------------------------------------------------------------

\end{document}